\documentclass[%
 reprint,
superscriptaddress,
nofootinbib,
 amsmath,amssymb,
 aps,
prx,
]{revtex4-2}

\usepackage[utf8]{inputenc}
\usepackage[english]{babel}
\usepackage{graphicx}
\graphicspath{{./figures/}}
\usepackage{dcolumn}
\usepackage{bm}
\usepackage{mathtools}
\usepackage{cancel}
\usepackage[caption=false,labelformat=parens,labelsep=quad,justification=raggedright]{subfig}

\usepackage{xcolor}

\renewcommand{\bf}[1]{\textbf{#1}}
\usepackage{ulem}
\usepackage{contour}

\contourlength{0.8pt}

\renewcommand{\emph}[1]{%
  \uline{\phantom{#1}}%
    \llap{\contour{white}{#1}}%
    }

\usepackage{xcolor}
\definecolor{purple}{rgb}{0.75,0.05,0.35}
\usepackage[colorlinks = true,
            linkcolor = blue,
            urlcolor  = blue,
            citecolor = blue,
            anchorcolor = blue,
    filecolor=blue]{hyperref}

\newcommand{\Eqref}[1]{Eq.~\eqref{#1}}
\newcommand{\secref}[1]{Sec.~\ref{#1}}
\def \ie {\textit{i.e.}\ }
\def \eg {\textit{e.g.}\ }
\def \cf {\textit{cf.}\ }
\def \testfnc {f}
\def\ts{\tilde{\omega}}

\def\eps{\varepsilon}

\def\cD{\mathcal{D}}

\def\cO{\mathcal{O}}
\def\cP{\mathcal{P}}

\def\cT{\mathcal{T}}
\def\adj{^{\dagger}}

\newcommand{\deltabar}{\delta\hspace*{-0.24em}\bar{}\hspace*{0.1em}\,}
\newcommand{\dbar}{\mathrm{d}\hspace*{-0.08em}\bar{}\hspace*{0.1em}}
\newcommand{\dd}[1]{\mathrm{d}{#1}\,}
\newcommand{\He}{\operatorname{He}}
\newcommand{\avg}[1]{\left\langle {#1} \right\rangle}
\newcommand{\yavg}[1]{\overline{#1}}

\def\deriv{\Delta}

\def\fptxy{\tau_{x_0,x_1}}

\newcommand{\fpt}[1]{\hat{p}\left(#1\right)}

\newcommand{\yfpt}[1]{\hat{p}\left(#1;[y]\right)}

\newcommand{\avgfpt}[1]{M\left(#1\right)}

\newcommand{\avgtrans}[1]{T\left(#1\right)}

\newcommand{\transcoeff}[2]{%
\hat{T}^{(#1)}\left(#2\right)
}%

\newcommand{\returncoeff}[2]{%
\hat{R}^{(#1)}\left(#2\right)
}%

\newcommand{\invreturncoeff}[2]{%
(\hat{R}^{-1})^{(#1)}\left(#2\right)
}%

\newcommand{\fullinvreturncoeff}[2]{%
    \IfEqCase{#1}{%
        {0}{\hat{R}^{-1}\left(#2\right)}%
        {1}{(\hat{R}^{-1})^{\prime}\left(#2\right)}%
        {2}{(\hat{R}^{-1})^{\prime \prime}\left(#2\right)}%
    }[(\hat{R}^{-1})^{(#1)}\left(#2\right)]%
}%

\newcommand{\corr}[1]{\hat{C}_2(#1)}

\def\target{\bar{x}_1}
\def\start{\bar{x}_0}

\def\LL{\mathcal{L}}

\usepackage{tikz}
\usetikzlibrary{decorations.pathmorphing}
\usetikzlibrary{decorations.markings}
\usetikzlibrary{arrows,shapes,snakes,automata,backgrounds,petri}
\usetikzlibrary{calc}
\usetikzlibrary{patterns}
\usetikzlibrary{decorations.text}
\tikzset{
	bareprop/.style={very thick,draw=red},
	ynoise/.style={very thick,dashed,draw=blue},
	bareproparrow/.style={very thick,draw=red, postaction={decorate},
decoration={markings,mark=at position .5 with {\arrow[draw=red]{>}}}},
}

\usepackage{tikz-3dplot}

\graphicspath{ {./figures/} }

\begin{document}


\title{First passage time distribution of active thermal particles in potentials}

\author{Benjamin Walter}
\affiliation{Department of Mathematics, Imperial College London, 180 Queen's Gate, SW7 2AZ London, United Kingdom}
\affiliation{Centre for Complexity \& Networks, Imperial College London, SW7 2AZ London, United Kingdom}
\affiliation{SISSA - International School for Advanced Studies, via Bonomea 265, 34135 Trieste, Italy}

\author{Gunnar Pruessner}
\affiliation{Department of Mathematics, Imperial College London, 180 Queen's Gate, SW7 2AZ London, United Kingdom}
\affiliation{Centre for Complexity \& Networks, Imperial College London, SW7 2AZ London, United Kingdom}
\author{ Guillaume Salbreux}
\affiliation{The Francis Crick Institute, 1 Midland Road, NW1 1AT London, United Kingdom}
\affiliation{ Department of Genetics and Evolution, University of Geneva, Quai Ernest-Ansermet 30, 1205 Geneva, Switzerland}
\date{\today}

\date{\today}

\begin{abstract}
We introduce a perturbative method to calculate all moments of the first-passage time distribution in stochastic one-dimensional processes which are subject 
to both white and coloured noise.  This class of non-Markovian processes is
at the centre of the study of 
thermal active matter, that is
self-propelled particles subject to diffusion. 
The perturbation theory about the Markov process
considers the effect
of self-propulsion to be small compared to that of thermal fluctuations. 
To illustrate our method, we apply it to the case of active thermal particles 
(i) in a harmonic trap
(ii) on a ring.
For both we calculate 
the first-order correction of the moment-generating function of first-passage times, and thus to all its moments. 
Our analytical results are compared to numerics.
\end{abstract}

\maketitle

\section{\label{sec:intro} Introduction and Main results}
\subsection{Introduction}
Understanding the statistical properties of first-passage times (FPT), the time a stochastic process takes to first reach a prescribed target, has enjoyed increased attention over the last two decades \cite{kampen_stochastic_2007, redner_guide_2001,metzler_first-passage_2014}, since it is a key characteristic of complex systems, such as chemical reactions \cite{szabo_first_1980}, polymer-synthesis \cite{sokolov_cyclization_2003}, intra-cellular events \cite{ghusinga_first-passage_2017}, neuronal activity \cite{taillefumier_phase_2013} or financial systems \cite{metzler_applications_2014}. Besides their dynamical information, FPTs are helpful to understand spatial properties of complex networks \cite{tejedor_global_2009}, extreme values of stochastic processes \cite{hartich_extreme_2019} and characteristic observables in out-of-equilibrium statistical physics \cite{bray_persistence_2013}.

\begin{figure}
    \centering
    \includegraphics[width=\columnwidth]{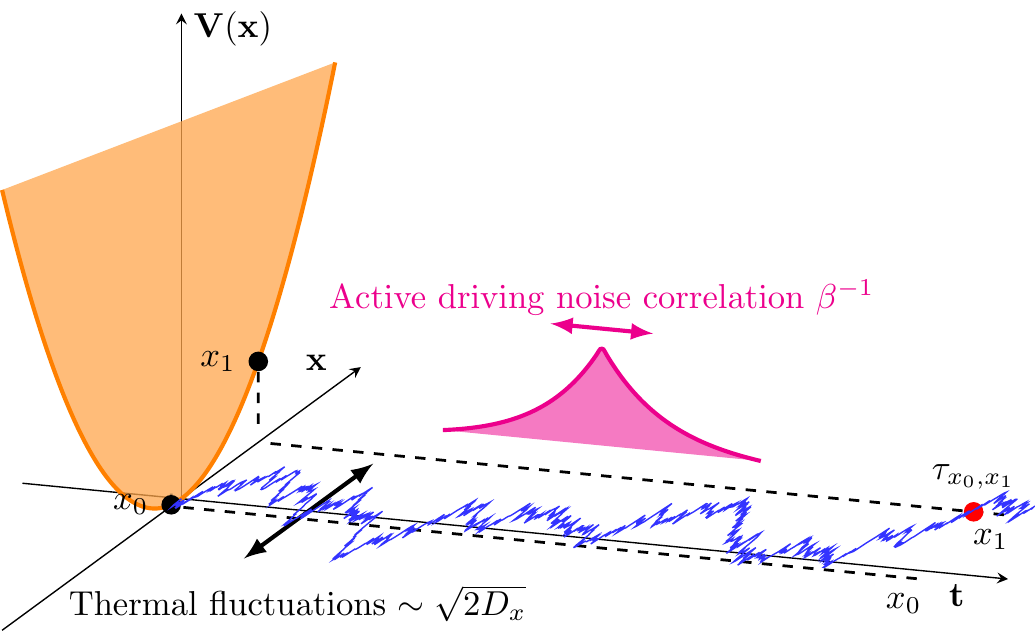}
    \caption{\textbf{A particle in a potential} (\textit{orange parabola}) \textbf{subject to both white and coloured noise} (see \Eqref{eq:driven_langevin}). While the white noise models a thermal environment whose timescale of correlation is negligibly small, the driving term models hidden degrees of freedom which are correlated over timescales comparable to those of the particle's stochastic dynamics. Those driving forces induce correlations (\textbf{pink correlation kernel}) in the particle's increments and therefore break its Markovianity. In this work, we study first-passage times $\fptxy$; the time such a random walker (\textbf{blue rough path}) takes to first reach $x_1$ starting from $x_0$ (\textbf{dashed lines}). }
    \label{fig:sketch_potential}
\end{figure}

For Markovian stochastic processes, the FPT has been studied for over a century, and often the full distribution of FPT, or their moment generating function, can be found in closed form \cite{erwin_schrodinger_zur_1915,pontryagin_appendix_1989,siegert_first_1951,darling_first_1953}. For non-Markovian processes, however, the problem of finding the FPT distribution is  much more difficult and often the focus has been on the mean first-passage time (MFPT) alone \cite{masoliver_first-passage_1986,kus_mean_1990,kus_mean_1991,ramrez-piscina_first-passage_1991,dykman_probability_1993,schimansky-geier_colored_1997,fox_mean_1988,hanggi_colored_1995,guerin_mean_2016}. The MFPT, however, can be insufficient to characterise complex dynamics, emphasizing the need for a more precise understanding of the full FPT distribution \cite{godec_first_2016,godec_universal_2016}. 

In this article, we address this challenge and compute the full moment-generating function of a class of non-Markovian stochastic processes perturbatively. In doing so, we obtain all moments of the distribution to the same order in the perturbative expansion. By numerical integration, we further recover the full FPT distribution With this method the full distribution is obtained systematically in the presence of correlated driving noise and white thermal noise for a wide range of settings, including a potential. The formulas we obtain order by order are exact, and the results we obtain for two systems, as an illustration, are in excellent agreement with numerical simulations. This allows for a detailed analysis of the qualitative changes of the FPT distribution  induced by correlations in stochastic forces. 

In this article, we consider the FPT problem in one dimension. 
We study the FPT of a particle at position $x_t$ placed in a potential $V(x)$, and subject to white \emph{thermal} noise $\xi_t$, $\avg{\xi_{t_0} \xi_{t_1}} = 2D_x \delta(t_1-t_0)$, modelling a surrounding heat bath at a temperature proportional to $D_x$ (via the fluctuation-dissipation theorem \cite{callen_irreversibility_1951,kubo_fluctuation-dissipation_1966}). Additionally, we assume that the particle is subject to a second stochastic force which may model either self-propulsion or hidden internal degrees of freedom. 
We refer to this force as \emph{active}, as we  equate ``activeness'' with the presence of coloured noise (\eg telegraphic noise or an Ornstein-Uhlenbeck noise), following \eg \cite{maggi_multidimensional_2015,fodor_how_2016,mandal_entropy_2017}.
The active noise $y_t$ is a stochastically independent stationary stochastic force with zero mean and a non-vanishing autocorrelation function $C_2(t_1-t_0) = \yavg{y_{t_0}y_{t_1}}$. 
Since the process includes two independent stochastic forces, we introduce $\avg{\bullet}$ to denote averages over $\xi_t$, and $\yavg{\bullet}$ to denote averages over $y_t$. 
The particle's position satisfies a Langevin equation \cite{kampen_stochastic_2007} of the type
\begin{align}
    \dot{x}_t = -V'(x_t) + \xi_t + \varepsilon y_t;
    \label{eq:driven_langevin}
\end{align}
where $\eps$ denotes a dimensionless coupling to the active term which is chosen small enough such that $\eps y_t \ll \xi_t$ in probability. We refer to the process described in \Eqref{eq:driven_langevin} as \emph{active thermal} process, as opposed to $(i)$ purely active (``athermal'', \cite{szamel_stochastic_2019}) processes and $(ii)$ purely thermal processes since $y_t$ breaks thermal equilibrium (\cite{luczka_non-markovian_2005}). Active thermal processes have recently been considered in \eg \cite{malakar_steady_2018, caprini_entropy_2019,sevilla_generalized_2019,dabelow_irreversibility_2019}. While the FPT problem in purely active matter such as ``run-and-tumble'' processes (\cite{cates_diffusive_2012}) has been studied in \cite{dhar2019runandtumble}.  
Understanding the FPT in active thermal matter is relevant to \eg neural activity \cite{Lindner2008}, transport in living cells \cite{Fodor2015}, and molecular motors \cite{Gabrys2003,pavliotis2005multiscale} where the influence of the thermal environment cannot necessarily be neglected.

\subsection{Main results}
\begin{figure*}
\subfloat[ \textbf{First-passage time distribution of active thermal Ornstein-Uhlenbeck Process (ATOU)} (\cf \Eqref{eq:ouou_SDE}) compared to numerical Laplace inversion of analytically obtained moment generating function (\cf theoretical result in \Eqref{eq:atou_mgf}) for various values of $\nu$ (solid lines). The plot marks indicate the distribution as sampled through Monte Carlo simulations with $5\times10^6$ runs. Simulation parameters are  $x_0 = 0, x_1 = 1, \alpha = 1, D_x = 1, D_y = 1, \beta = \frac{1}{2}$ while $\eps$ is tuned to fix $\nu = D_y \beta \eps^2/(D_x \alpha)$ to values as indicated in legend. The \textbf{inset} shows rescaled deviations to the undriven first-passage time distribution (plot marks, \cf \Eqref{eq:num_estimate_f1}) as compared to the first-order correction (solid lines, $\nu=0.1$ omitted). Higher-order corrections appear for growing values of $\nu$ (cf.~Fig.~\ref{fig:ouou_m1}). See Sec.~\ref{subsubsec:numerical_validation_ouou} for discussion.] 
  {\includegraphics[width=\columnwidth]{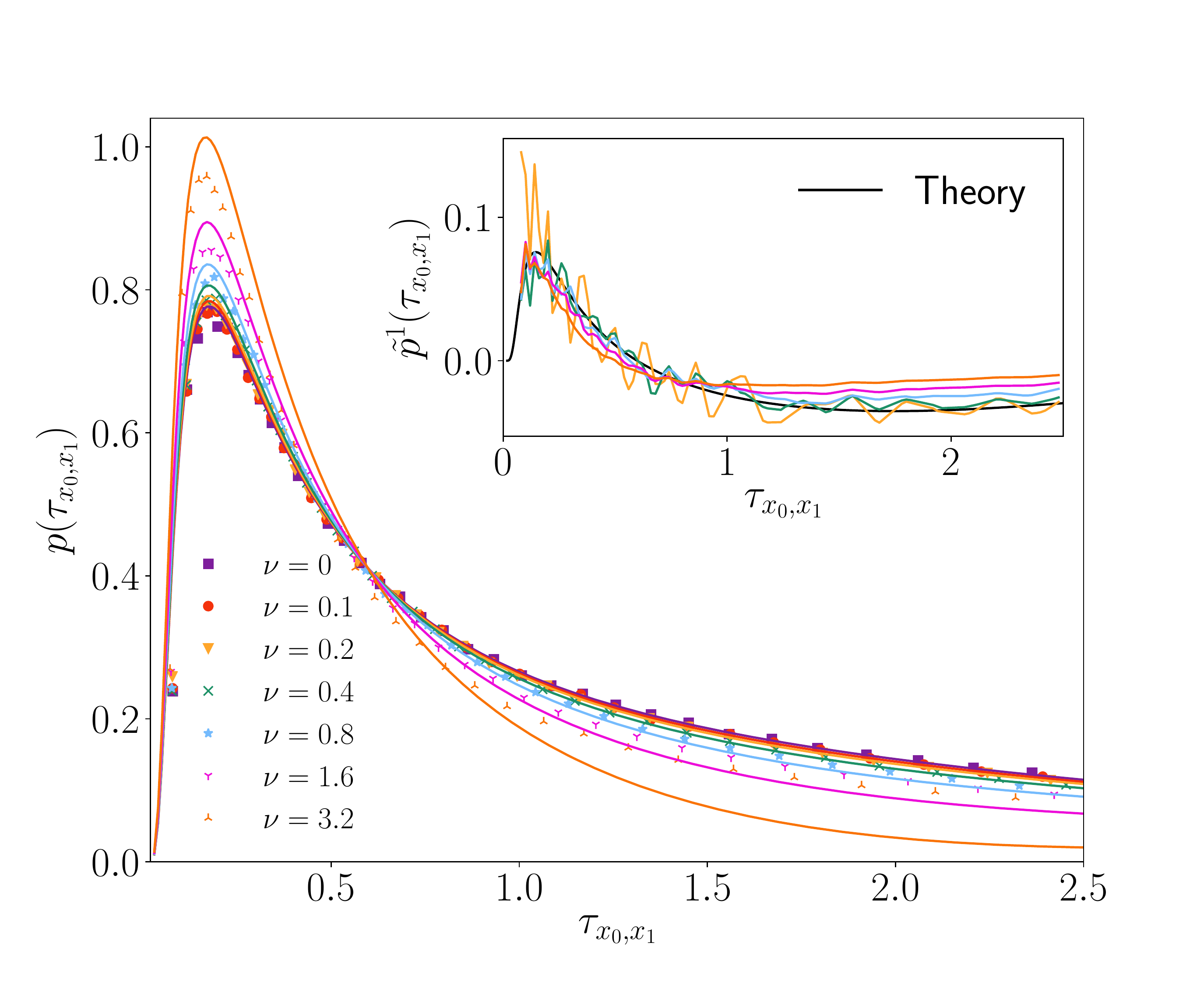} \label{fig:ouou_pdf}}
\hfill
 \subfloat[
 \textbf{First-passage time distribution of active thermal Brownian Motion (ATBM)} (\cf \Eqref{eq:bmou_sde}) compared to numerical Laplace inversion of analytically obtained moment generating function (\cf theoretical result in \Eqref{eq:bmou_full_mgf}) for various values of $\nu$ (solid lines). The plot marks indicate the distribution as sampled through Monte Carlo simulations with $10^6$ runs. Simulation parameters are  $x_0 = 0, x_1 = \pi, r = 1, D_x = 1, D_y = 1, \beta = \frac{1}{2}$ while $\eps$ is tuned to fix $\nu = D_y r^2 \beta \eps^2/D_x^2 $ to values as indicated in legend. The \textbf{inset} shows rescaled deviations to the undriven first-passage time distribution (\cf \Eqref{eq:num_estimate_f1}) as compared to the first-order correction (solid lines, $\nu=0.1$ omitted). Higher-order corrections appear for growing values of $\nu$ (cf.~Fig.~\ref{fig:mgf_bmou}). See Sec.~\ref{subsec:atbmnumerics} for discussion.]{\includegraphics[width=\columnwidth]{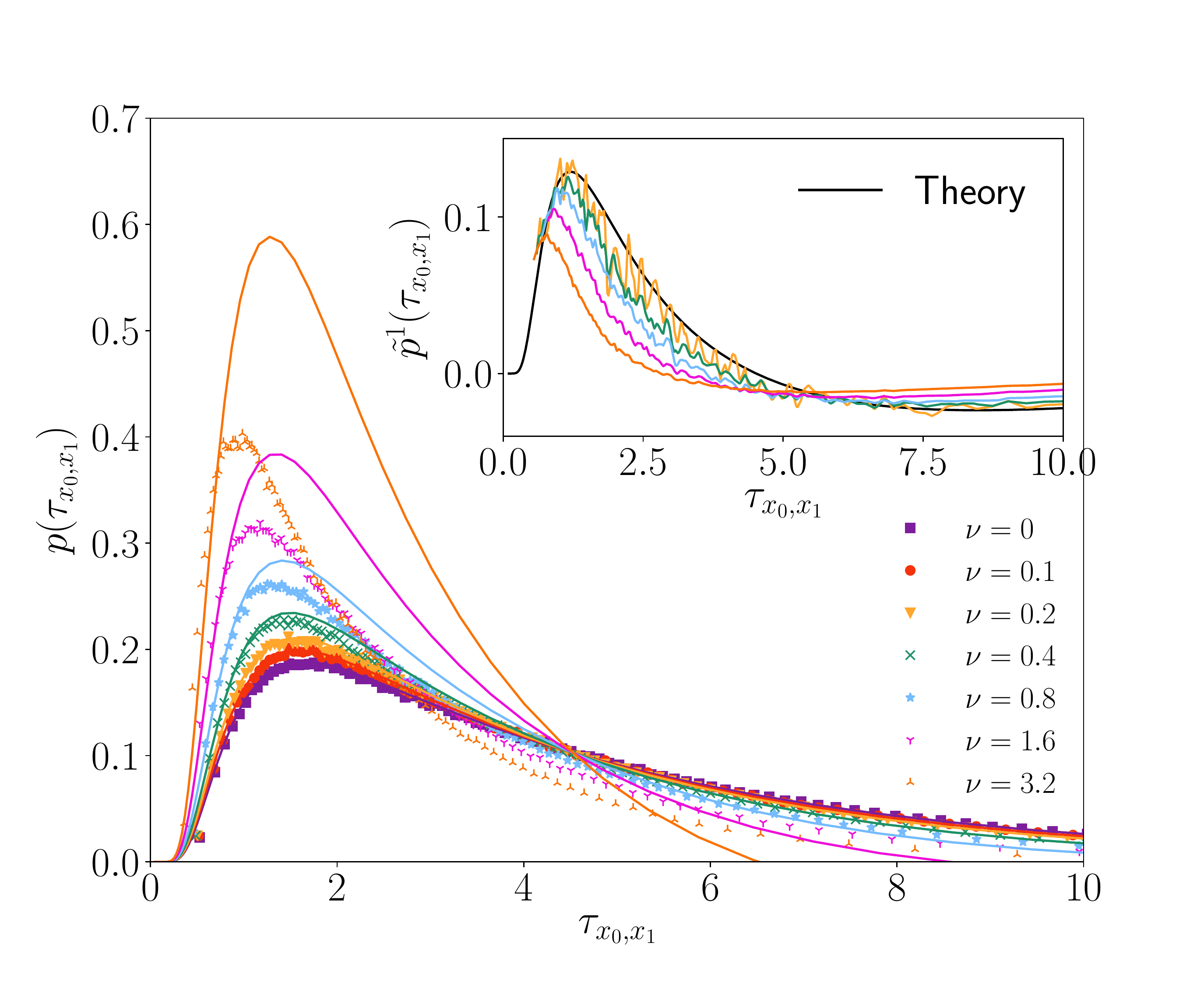}\label{fig:bmou_pdf}} 
  \caption{First order corrections to the probability distribution of first-passage times as found by the framework presented in this work for two example processes.}
\end{figure*}

In the following we denote  $\fptxy$ the first-passage time of $x_t$ defined as
\begin{align}
    \fptxy \coloneqq \inf_{t_1>t_0} \left\{ t: x_{ t_1} = x_1 | x_{t_0} = x_0  \right\},
\end{align}
and $p_{x_0, x_1}(t_0,t_1)=p_{x_0, x_1} (t_1-t_0)$ the probability distribution of $\fptxy$, where time-invariance of the $y_t$-averaged process ensures that $p_{x_0, x_1}$ depends only on the time difference $t_1-t_0$.
The central result of the present work concerns the moment-generating function of first-passage times of $x_t$, 
\begin{multline}
   M_{x_0, x_1}(s)= \avg{\yavg{e^{-s \fptxy}}} \\
    = 1 - s \avg{\yavg{\fptxy}} + \frac{s^2}{2}\avg{\yavg{\left(\fptxy\right)^2}} + ...
\end{multline}
for $s\geq 0$. We note that $M_{x_0, x_1} (s)$ is the Laplace transform of the probability distribution $p_{x_0, x_1}(t)$, and the restriction to the negative imaginary axis of the complex function $\hat{p}_{x_0,x_1}(z)=\int_0^{\infty} \dd{t} p_{x_0,x_1}(t) e^{-i z t}$. Because $p_{x_0, x_1} (t)$ is non-zero only for $t\geq 0$, and $\int_0^{\infty} \dd t p_{x_0, x_1} (t)=1$, the function $\hat{p}_{x_0, x_1} (z)$ is bounded, and by the Paley-Wiener theorem holomorphic, on the lower half complex plane \cite{Hoermander1963}. Besides, $\hat{p}_{x_0, x_1} (z)$ converges to the Fourier transform of $p_{x_0, x_1} (t)$ on the real axis (defined in Eq. \ref{eq:Fourier_transform_functional}). Therefore in the following we will use alternatively the Fourier transform and Laplace transform for our calculations whenever convenient, and assume that conditions are met such that a simple rotation $\omega=-is$ allows to go from one transform of $p_{x_0, x_1}(t)$ to the other.

We now assume that the moment generating function $M_{x_0, x_1}(s)$ has an expansion of the form
\begin{align}
    M_{x_0, x_1} (s)= M^0_{x_0,x_1} (s) + \nu M^1_{x_0,x_1}(s) + \cO(\nu^2).
    \label{eq:general_f_structure}
\end{align}
where $M^0$ and $M^1$ are the coefficients of expansion of $M_{x_0, x_1} (s)$ in a dimensionless parameter $\nu$ of order $\cO(\eps^2)$, around $\nu = 0$. The case $\eps = 0$ corresponds to a purely thermal process, and the corresponding moment-generating function $M^0$, can be found by classical methods such as the Darling-Siegert method which is discussed in the next subsection. Here we assume that around this state $M_{x_0, x_1} (s)$ is analytic in $\nu$.

 The first-order contribution, $M^1$, requires some deeper analysis. Much of what follows is dedicated to the calculation of $M^1$, which to our knowledge is new in the literature. 
In principle, the method we present here is capable of calculating coefficients $M^2, M^3,...$ of arbitrarily high order of $\nu$ for arbitrarily coloured noise  (cf. \Eqref{eq:colerator}) as long as the autocorrelations can be integrated suitably. Furthermore the potentials  $V(x)$ are arbitrary, as long as an associated differential operator can be diagonalised (Sec.~\ref{sec:results} and \cite{bakry_analysis_2013}), otherwise it needs to be treated perturbatively as well.

We further illustrate our framework by explicitly computing $M^0$ and $M^1$ for two cases, each of which are additionally driven by coloured Gaussian noise, \ie $y_t$ is Gaussian and has correlator $\yavg{y_{t_0}y_{t_1}} = D_y\beta e^{-\beta|t_1-t_0|}$ with some diffusivity $D_y$ and correlation time $\beta^{-1}$. In the first case, the particle is placed in a harmonic potential, $V(x) = \frac{\alpha}{2}x^2$. This particular model has been studied in, \eg, \cite{dabelow_irreversibility_2019}. We refer to this model as active thermal Ornstein Uhlenbeck process (ATOU). While $M^0$ (see \Eqref{eq:OUOU_M0}) has been long known, $M^1$ (see \Eqref{eq:OUOU_M1}) is  a new result.  By numerically computing its inverse Laplace transform, we obtain the full probability distribution of first-passage times to first order in $\nu$. In Fig.~\ref{fig:ouou_pdf} we compare this to the numerically obtained probability density function. 
Secondly, we calculate $M^0$ and $M^1$ for the case of a Brownian Motion on a ring of radius $r$ driven by coloured Gaussian noise which we refer to as active thermal Brownian Motion (ATBM). Again, through numerically computing its inverse Laplace transform we obtain the corrected probability distribution of first-passage times which is compared to numerics in Fig.~\ref{fig:bmou_pdf}. 

Our method is \emph{systematic} since it allows its user to calculate in principle corrections to arbitrary order, and it is \emph{controlled} in the sense that the error can be made arbitrarily small. Further \emph{all moments} are available at once and to equal perturbative order. It is also valid for arbitrary noise colours $\beta^{-1}$. 

The paper is structured as follows. In Sec.~\ref{sec:methods} we give detailed account of how to calculate $M_{x_0,x_1}(s)$ for small $\nu$. First, we reproduce the Darling-Siegert argument in the equilibrium case ($\nu = \eps = 0$). Next, we introduce a perturbative version of the Darling-Siegert equation. Then, we obtain, as an intermediate result, a formula for $M_{x_0,x_1}(s)$ which is still a functional of the colored noise $y_t$. In the last step, we need to average over the stationary distribution of $y_t$ to arrive at the explicit formula \Eqref{eq:fpt_correction_general} which is the main result of our work. In the subsequent section \ref{sec:results}, we calculate all quantities required for the case of a harmonic potential and a Brownian motion with periodic boundary conditions and arrive at the first-order correction to the moment-generating function of first-passage times \Eqref{eq:OUOU_M1}. Section \ref{sec:discussion} concludes with a discussion of our findings.

\section{\label{sec:methods} Perturbation Theory}

As outlined above, in this work we present a way to calculate the moment-generating function of first-passage times of stochastic processes which are close to an equilibrium state. The underlying assumption is that the moment-generating function varies smoothly as $\varepsilon$, the coupling to the self-propelling force, is switched on. 
The moment-generating function of the  equilibrium version of the process ($\varepsilon = 0$) is assumed to be known in closed form, as is for instance justified for the Ornstein-Uhlenbeck process or Brownian Motion (\cite{bakry_analysis_2013}). This exact form is then corrected by terms in the spirit of a perturbative expansion which is controlled by powers of a dimensionless parameter describing the distance to equilibrium. 
First, we revise the arguments given by Darling and Siegert for the equilibrium case (\cite{siegert_first_1951,darling_first_1953}). Next, we outline our perturbative approach to the active case.

Before going further, we introduce some notations (see appendix \ref{appendix:notations} for a list). The transition probability density of progressing from $x_0$ at $t_0$  to $x_1$ at $t_1$ is denoted by 
\begin{align}
    T_{x_0, x_1}(t_0, t_1) = T(t_0,t_1)
\end{align}
where the arguments $x_0$ and $x_1$ will later be dropped wherever confusion can be avoided, for the sake of easier notation. 
Analogously the return probability at $x_1$, $T_{x_1,x_1}( t_0,t_1)$ is denoted by 
\begin{align}
   R_{x_1}( t_0,t_1) = R(t_0,t_1)= T_{x_1,x_1}( t_0,t_1)
\end{align}
In the following, we denote the Fourier transform of a function $f(t)$ by
\begin{align}
\label{eq:Fourier_transform_functional}
\hat{f}(\omega)=\int_{-\infty}^{\infty} \dd{t} e^{-i \omega t} f(t)~,
\end{align}
with inverse
\begin{align}
\label{eq:Fourier_transform_inverse_functional}
f(t)=\int_{-\infty}^{\infty} \dbar\omega\, e^{i \omega t} f(\omega),
\end{align}
 where $\dbar{\omega} = \frac{\dd{\omega}}{2\pi }$.

In the following, we consider functions $f(t_0,t_1)$ which depend on the difference $t_1-t_0$ only, i.e. $f(t_0,t_1)=f(t_1-t_0)$. We refer to functions of this type  as \emph{diagonal}. The Fourier transform in two variables of these functions read
\begin{align}
  \hat{f}(\omega_0, \omega_1) &=\int_{-\infty}^{\infty} \dd{t_0}\int_{-\infty}^{\infty}\dd{t_1}  e^{- i \omega_0 t_0 - i  \omega _1 t_1} f(t_0,t_1)\nonumber\\
&  =\hat{ f}(\omega_1) \deltabar(\omega_0 + \omega_1) ~,  \label{eq:diagonal_function}
\end{align}
where $\deltabar(\omega) =2\pi \delta(\omega)$.

\subsubsection{Equilibrium case: The Darling-Siegert solution}

We here consider the equilibrium case of \Eqref{eq:driven_langevin} defined by setting $\eps=0$. As $x_t$ is Markovian, the functions $p$ and $T$ satisfy the following renewal equation: 
\begin{align}
\label{eq:conditional_argument}
	T_{x_0, x_1}(t_0,t_1)= 
	\int_{t_0}^{t_1} \dd{t'} p_{x_0, x_2}(t_0,t')  T_{x_2, x_1}(t',t_1)
\end{align}
for all $x_2 \in (x_0,x_1]$. 
 Applying a Fourier transform to Eq.~\eqref{eq:conditional_argument}, the time-homogeneity of both $T_{x_0, x_1}(t_0,t_1) $ and $p(t_0,t_1) $ translates into diagonality in frequencies and turns the convolution into a product, such that the result can be stated at the level of the amplitudes alone. Rearranging the terms and choosing $x_2=x_1$  results in 
\begin{align}
	\hat{p}(\omega) = \frac{\hat{T}(\omega)}{\hat{R}(\omega)}~.
	\label{eq:FPTMGFclassic}
\end{align}
Since the Fourier transform of a probability density equals its characteristic function, \Eqref{eq:FPTMGFclassic} recovers all moments of the first passage time provided $\hat{T}$ is known. Further, setting $\omega = -i s$ for some $s \in \mathbb{R}_+$ turns the Fourier transforms into Laplace transforms and the characteristic function into the moment generating function. This recovers the Darling-Siegert equation in its original form in which the Laplace transform of transition and return probabilities is linked to the moment-generating function of first-passage times.
\subsubsection{Out-of-equilibrium: A perturbative approach}
The argument made by Darling and Siegert breaks down when $\eps \neq 0$: indeed when averaging over the driving noise $y(t)$, the renewal equation \Eqref{eq:conditional_argument} no longer is true. The approach we take in this paper, consists of three steps:
\begin{enumerate}
    \item Fix a particular realisation $y(t)$, and expand transition and return probabilities of $x_t$ as functional expansion around $y= 0$ of the form
    \begin{multline}
    \label{eq:functional_expansion_initial_exposition}
   \hat{T}(\omega_0,\omega_1,[\hat{y}]) 
    = \sum_{n=0}^{\infty} \frac{\eps^n}{n!} \int \dbar{\ts_1}...\dbar{\ts_n} \deltabar\left(\omega_0 + \omega_1 + \sum_{i=1}^n \ts_i\right)\\\times \transcoeff{n}{\omega_1,\ts_1,...,\ts_n}\hat{y}(-\ts_1)... \hat{y}(-\ts_n) ~.
\end{multline}
    \item As long as $y(t)$ is fixed, the process, when understood as conditioned on this particular driving, satisfies a renewal equation of the type \Eqref{eq:conditional_argument}. Inserting the perturbative transition and return probabilities from the previous step, gives a perturbative series for the first-passage time density $\yfpt{\omega_0, \omega_1}$ of $x_t$ conditioned on a particular $y_t$.
    \item Averaging over the ensemble of driving noises. For simplicity, we here assume that the correlation function of the driving noise is given by 
    \begin{align}
        \yavg{y_{t_0}y_{t_1}} = D_y\beta e^{-\beta|t_1 - t_0|}~,
    \end{align}
    where $\beta$ is the inverse correlation time. Generally, when computing the term of order $\eps^n$, the first $n$ moments of $y_t$ need to be known.
\end{enumerate}
This procedure leads to the central result of this work: the moment-generating function of first-passage times to second order in $\eps$ reads
\begin{widetext}
\begin{align}
	 \hat{p}(\omega) =&\frac{
		\transcoeff{0}{\omega}}
		{
	\returncoeff{0}{\omega}
	} 
	+ \frac{\varepsilon^2 D_y\beta}{2} \left[\frac{ \transcoeff{2}{\omega;i\beta,-i\beta}}{\returncoeff{0}{\omega}} 
		- 2 \frac{\transcoeff{1}{\omega-i\beta,i\beta} \returncoeff{1}{\omega,i\beta}}{\returncoeff{0}{\omega-i\beta} \returncoeff{0}{\omega}} \right. 
		\\
		& \qquad \qquad \qquad
  \left. 
		+2 \frac{\transcoeff{0}{\omega} \returncoeff{1}{\omega-i\beta,i\beta} \returncoeff{1}{\omega,i\beta}}{\left( \returncoeff{0}{\omega} \right)^2 \returncoeff{0}{\omega-i\beta}} 
		- \frac{\transcoeff{0}{\omega} \returncoeff{2}{\omega;i\beta,-i\beta}}{ \left( \returncoeff{0}{\omega} \right)^2} 
\right] + \cO(\eps^4).\nonumber
\end{align}
\end{widetext}

In the next sections, we derive this relation in more details.
\subsection{Perturbative Darling-Siegert equation}
\label{seq:perturbative_darling_siegert}

\subsubsection{Expression for the first-passage time distribution}
By imposing an additional driving noise $y_t$, the transition probability and FPT probability density of $x_t$  depend on a particular  realisation of $y_t$. Accordingly, we introduce the  transition probability density $T(t_0,t_1,[y]) $ and FPT probability density $p(t_0,t_1,[y]) $ as the densities of the process $x_t$ conditioned on  $y$ given. The conditional densities are explicitly dependent on $t_0$ and $t_1$ rather then their difference $t_1-t_0$ because $y_t$ is an explicit function of time.

For $y$ fixed, the process remains Markovian and therefore \Eqref{eq:conditional_argument} still applies and gives rise to:
\begin{multline}
\label{eq:conditional_argument_2}
	T_{x_0,x_1}(t_0,t_1;[y])= 
	\int_{-\infty}^{\infty} \dd{t'} p_{x_0,x_1}(t_0,t';[y]) \\
	\times R_{x_1}(t',t_1;[y])~,
\end{multline}
where the dependency of the functions on the spatial values $x_0$, $x_1$ has been made explicit for clarity, and we have used the fact that $p(t_0, t';[y])$ vanishes for $t'<t_0$ and $R(t',t_1)$ vanishes for $t'>t_1$ to integrate over the full real axis.

It is no longer possible to directly invert this equation in Fourier space to solve for $p$, as done in \Eqref{eq:FPTMGFclassic}, since neither terms in the integral are diagonal, \ie they depend explicitly on both $t_0, t'$ and $t', t_1$. However, introducing the inverse functional $R^{-1}$ which is defined by the implicit equation:
\begin{align}
\label{eq:def_inverse_R}
\int_{-\infty}^{\infty} \dd{t} R^{-1}(t_0, t;[y]) R(t, t_1;[y]) =\delta(t_1-t_0)~,
\end{align}
the renewal equation can be formally solved by the relation:
\begin{multline}
\label{eq:main_eq_to_solve}
p_{x_0,x_1}(t_0, t_1;[y])\\
=\int_{-\infty}^{\infty} \dd{t} T_{x_0,x_1}(t_0, t;[y])  R^{-1}_{x_1}(t,t_1;[y])~.
\end{multline}

Our approach is then to perform a functional expansion of the quantities involved in \Eqref{eq:main_eq_to_solve} in the function $y$, around the Markovian case $y=0$. 

\subsubsection{Functional expansion of the transition and return probability densities}

We start by performing a Taylor expansion of the transition probability $T(t_0, t_1)$:
\begin{multline}
  T(t_0,t_1;[y]) 
  =\sum_{n=0}^{\infty}\frac{1}{n!} \int  \dd{\tilde{t}_1}...\dd{\tilde{t}_n}  \\
   \frac{\delta^n T }{\delta y(\tilde{t}_1)...\delta y(\tilde{t}_n)}(t_0,t_1;[0]) y(\tilde{t}_1)...y(\tilde{t}_n)~.
\end{multline}
We now argue that this functional expansion can be rewritten in a simpler form, using the time-invariance properties of the dynamic equation \eqref{eq:driven_langevin}. We introduce the time-shifting operator which transforms a function $y_t$ into a time-shifted function $\tilde{y}_t$, according to:
\begin{align}
S_{t_0}[y](t)=y_{t+t_0}\equiv\tilde{y}_t
\end{align}
and we will use the fact that:
\begin{align}
\label{eq:time_shifting_invariance}
  T(t_0,t_1;[y]) =  T(t_0+\tau,t_1+\tau;[S_{-\tau}[y]])
\end{align}
for any time-shift $\tau$. This relation results from the invariance by time shifting of \Eqref{eq:driven_langevin}, when the function $y_t$ is shifted in time accordingly. As a result we introduce the shifted function $T_0$, defined by;
\begin{align}
\label{eq:definition_T_0}
T_0(\tau; [\tilde{y}])=T(0, \tau; [\tilde{y}])
\end{align}
such that \Eqref{eq:time_shifting_invariance} gives, choosing $\tau=-t_0$:
\begin{align}
\label{eq:main_identity_time_shifting}
T(t_0,t_1;[y])=T_0(t_1-t_0; [S_{t_0}[y]])~.
\end{align}
Using \Eqref{eq:main_identity_time_shifting}, the functional derivatives of $T$ can be related to the functional derivatives of $T_0$ through
\begin{multline}
\frac{\delta^n T}{\delta y(\tilde{t}_1)...\delta y(\tilde{t}_n)} (t_0,t_1;[y]) =\\
\frac{\delta^n T_0}{\delta \tilde{y}(\tilde{t}_1-t_0)...\delta \tilde{y}(\tilde{t}_n-t_0)}  (t_1-t_0;[S_{t_0}[y]])~,
  \end{multline}
and the functional expansion of $T$ can now be written
\begin{multline}
\label{eq:intermediate_step_functional_expansion}
  T(t_0,t_1;[y]) =\sum_{n=0}^{\infty}\frac{1}{n!} \int  \dd{\tilde{t}_1}...\dd{\tilde{t}_n}\\
  \times  \frac{\delta^n T_0 }{\delta \tilde{y}(\tilde{t}_1-t_0)...\delta \tilde{y}(\tilde{t}_n-t_0)}(t_1-t_0;[0])y(\tilde{t}_1)...y(\tilde{t}_n)~,
\end{multline}
where we have used that $S_{t_0}[y=0]= [0]$.
So far, the transitional probability $T(t_0,t_1;[y])$ has been expressed as a functional of $y$ in the time-domain. In order to consider the transition probability as a functional of the Fourier-transformed driving noise $\hat{y}$, we introduce $\hat{T}(\omega_0,\omega_1;[\hat{y}])$, defined by taking the Fourier transform of $T(t_0, t_1, [y])$ with respect to $t_0$, $t_1$, and taking $y$ equal to the inverse Fourier transform of $\hat{y}$ in the definition of $T$:
\begin{align}
\label{eq:def_hatT}
\hat{T}(\omega_0,\omega_1;[\hat{y}])&=\int \dd{t_0}   \dd{t_1} T(t_0 ,t_1, [y]) e^{-i (\omega_0 t_0+\omega_1 t_1)}~,
 \end{align}
 and similarly:
 \begin{align}
\hat{T}_0(\omega ;[\hat{y}])&=\int \dd{\tau} T_0(\tau, [y]) e^{-i \omega \tau}~.
 \end{align}
The functional derivatives of $T_0(\tau;[y])$ and $\hat{T}_0(\omega;[\hat{y}])$ are related by
\begin{align}
 \frac{\delta^n T_0(\tau;[y])}{\delta y(\tilde{t}_1)...\delta y(\tilde{t}_n)} &=\int \dbar{\omega}\dd{\tilde{\omega}_1}...\dd{\tilde{\omega}_n} \frac{\delta ^n \hat{T}_0(\omega;[\hat{y}])}{\delta \hat{y}(\tilde{\omega}_1)...\delta \hat{y}(\tilde{\omega}_n)}\nonumber\\
& \qquad \times e^{-i \tilde{\omega}_1 \tilde{t}_1}...e^{-i\tilde{\omega}_n \tilde{t}_n} e^{i \omega \tau}~.
\end{align}
As a result, combining  \Eqref{eq:intermediate_step_functional_expansion} and \Eqref{eq:def_hatT}, we obtain the functional expansion:
    \begin{multline}
    \hat{T}(\omega_0,\omega_1,[\hat{y}])
    \\
    = \sum_{n=0}^{\infty}\frac{1}{n!} \int \dd{\ts_1}...\dd {\ts_n}\deltabar\left(\omega_0 + \omega_1 - \sum_{i=1}^n \ts_i\right) \\ \times \frac{\delta^n \hat{T}_0 }{\delta \hat{y}(\tilde{\omega}_1)...\delta \hat{y}(\tilde{\omega}_n)} (\omega_1;[0]) \hat{y}(\ts_1)... \hat{y}(\ts_n) 
    	\label{eq:rho_functional_expansion}~.
\end{multline}

The perturbative expansion of the return probability $R$ is simply obtained from by letting $x_0 \to x_1$ in the expansion of $T$ and has therefore the same structure, using $R_0(\tau; [\tilde{y}])=R(0, \tau; [\tilde{y}])$. To ease the notation, we now introduce a shorter notation for the set of functions:
\begin{align}
\label{eq:def_t_expansion}
T^{(n)}(\omega_1,\tilde{\omega}_1,...,\tilde{\omega}_n)=
\frac{(2\pi)^n}{\eps^n}\frac{\delta^n \hat{T}_0}{\delta \hat{y}(-\tilde{\omega}_1)...\delta \hat{y}(-\tilde{\omega}_n)} (\omega_1;[0]) \\
R^{(n)}(\omega_1,\tilde{\omega}_1,...,\tilde{\omega}_n)=
\frac{(2\pi)^n}{\eps^n}\frac{\delta^n \hat{R}_0}{\delta \hat{y}(-\tilde{\omega}_1)...\delta \hat{y}(-\tilde{\omega}_n)}(\omega_1;[0]) 
\label{eq:def_r_expansion}
\end{align}
where the factor $1/\eps^n$ has been introduced for convenience, to make the expansion in small $\eps$ explicit in \Eqref{eq:rho_functional_expansion}. We then arrive at the following form:
    \begin{multline}
    \label{eq:functional_expansion_second_exposition}
   \hat{T}(\omega_0,\omega_1,[\hat{y}]) 
    = \sum_{n=0}^{\infty} \frac{\eps^n}{n!} \int \dbar{\ts_1}...\dbar{\ts_n} \deltabar\left(\omega_0 + \omega_1 + \sum_{i=1}^n \ts_i\right)\\\times \transcoeff{n}{\omega_1,\ts_1,...,\ts_n}\hat{y}(-\ts_1)... \hat{y}(-\ts_n)\\
       \hat{R}(\omega_0,\omega_1,[\hat{y}]) 
    = \sum_{n=0}^{\infty} \frac{\eps^n}{n!} \int \dbar{\ts_1}...\dbar{\ts_n} \deltabar\left(\omega_0 + \omega_1 + \sum_{i=1}^n \ts_i\right)\\\times \returncoeff{n}{\omega_1,\ts_1,...,\ts_n}\hat{y}(-\ts_1)... \hat{y}(-\ts_n) ~,
\end{multline}
where we have made a choice of signs of $\tilde{\omega}_1,..\tilde{\omega}_n$ for convenience.
\subsubsection{Functional expansion of the inverse of the return probability density}

We now turn to the expansion of $R^{-1}$ in \Eqref{eq:main_eq_to_solve}. As $T$ and $R$, $R^{-1}$ satisfies the time-shift invariance relation \eqref{eq:time_shifting_invariance}. This can be seen by time shifting \Eqref{eq:def_inverse_R} by the time $\tau$:
\begin{multline}
\int_{-\infty}^{\infty} \dd{t} R^{-1}(t_0+\tau, t+\tau;S_{-\tau}[y])\\
\times R(t+\tau, t_1+\tau;S_{-\tau}[y]) =\delta(t_1-t_0)
\end{multline}
and, since $R(t+\tau, t_1+\tau;S_{-\tau}[y]) =R(t,t_1,[y])$, one also obtains $R^{-1}(t_0+\tau, t+\tau;S_{-\tau}[y])=R^{-1}(t_0, t;[y])$ since \Eqref{eq:def_inverse_R} defines $R^{-1}$. The analysis of the previous subsection still applies and one obtains the expansion
\begin{align}
\label{eq:expansion_Rminus1}
\hat{R}^{-1}(\omega_0,\omega_1,[\hat{y}])= \sum_{n=0}^{\infty} \frac{\eps^n}{n!} \int \dbar{\ts_1}...\dbar{\ts_n} \deltabar\left(\omega_0 + \omega_1 + \sum_{i=1}^n \ts_i\right)\\
\times (\hat{R}^{-1})^{(n)}(\omega_1,\ts_1,...,\ts_n)\hat{y}(-\ts_1)... \hat{y}(-\ts_n) ~. \nonumber
\end{align}
\begin{widetext}
We then obtain by applying a Fourier transform to \Eqref{eq:def_inverse_R}:
\begin{align}
\int \dbar \omega \hat{R}^{-1}(\omega_0, \omega;[\hat{y}]) \hat{R}(-\omega,\omega_1;[\hat{y}])= \deltabar(\omega_0 +\omega_1) ~.
\end{align}
This relation allows to relate the functions  $(R^{-1})^{(n)}$ to the functions $R^{(n)}$, by plugging Eqs. \eqref{eq:functional_expansion_second_exposition} and \eqref{eq:expansion_Rminus1} and identifying order by order in $\eps$. Limiting ourselves up to order $2$, we obtain:
\begin{align}
(\hat{R}^{-1})^{(0)}(\omega) &= \frac{1}{\hat{R}^{(0)}(\omega)} 
	\label{eq:inverse_rho_zero}\\
(\hat{R}^{-1})^{(1)}(\omega, \ts_1)& =  -\frac{\returncoeff{1}{\omega, \ts_1}}{\returncoeff{0}{\omega}\returncoeff{0}{\omega+\ts_1}}
\label{eq:inverse_rho_first}\\
	\label{eq:inverse_rho_second}
	\invreturncoeff{2}{ \omega, \ts_1, \ts_2} 
	&= \frac{1}{\returncoeff{0}{\omega+\ts_1+\ts_2}\returncoeff{0}{\omega}}\Bigg\{
	\frac{2\returncoeff{1}{ \omega+\ts_2, \ts_1} \returncoeff{1}{\omega, \ts_2}}{ \returncoeff{0}{  \omega+\ts_2} }  -\returncoeff{2}{\omega, \ts_1, \ts_2} \Bigg\}.
\end{align}

\subsubsection{Second-order expansion of the first passage time distribution}
Equipped with these expansions, one now can expand the first-passage time density expression \eqref{eq:main_eq_to_solve} to obtain a functional expansion of the first-passage density in $\eps$, involving the functions $\hat{T}^{(n)}$ and $\hat{R}^{(n)}$ which are simpler to calculate. We first need to write down the Fourier-transformed version of \Eqref{eq:main_eq_to_solve}:
\begin{align}
\label{eq:main_eq_to_solve_Fourier}
\hat{p}(\omega_0, \omega_1;[\hat{y}])=\int_{-\infty}^{\infty} \dbar\omega  \, \hat{T}(\omega_0, \omega;[\hat{y}])\, \hat{R}^{-1}(-\omega,\omega_1;[\hat{y}])~.
\end{align}
where the dependency on $x_0$, $x_1$ is here implicit.
Performing an expansion of this relation in $\eps$, the result reads to second order:
\begin{align}
	\label{eq:fpt_expansion_in_y}
&\hat{p}(\omega_0,\omega_1;[\hat{y}]) = \frac{\hat{T}^{(0)}(\omega_1)}{\returncoeff{0}{\omega_1}}\deltabar(\omega_0 +\omega_1) 
+ \eps \left[ -  \frac{\transcoeff{0}{-\omega_0}\returncoeff{1}{\omega_1,-\omega_0-\omega_1}}{\returncoeff{0}{-\omega_0}\returncoeff{0}{\omega_1}}
+ \transcoeff{1}{\omega_1,-\omega_0-\omega_1}\frac{1}{\returncoeff{0}{\omega_1}} \right] \hat{y}(\omega_0+\omega_1) 
\nonumber \\
& + \frac{1}{2}\eps^2\int \dbar{\ts} \bigg[\frac{\transcoeff{0}{-\omega_0}}{\returncoeff{0}{-\omega_0}\returncoeff{0}{\omega_1}}\bigg\lbrace   \frac{2\returncoeff{1}{-\ts-\omega_0;\ts}\returncoeff{1}{\omega_1,-\omega_0-\omega_1-\ts}}{\returncoeff{0}{-\ts-\omega_0}} -\returncoeff{2}{\omega_1,\ts,-\omega_0-\omega_1-\ts}\bigg\rbrace 
\nonumber \\
& + \frac{\transcoeff{2}{\omega_1,\ts,-\omega_0-\omega_1-\ts}}{\returncoeff{0}{\omega_1}}  -\frac{2\transcoeff{1}{-\ts-\omega_0;\ts}\returncoeff{1}{\omega_1,-\omega_0-\omega_1-\ts}}{\returncoeff{0}{\omega_1}\returncoeff{0}{-\ts-\omega_0}}\bigg]\hat{y}(-\ts)\hat{y}(\omega_0+\omega_1+\ts) + ...
\end{align}
\end{widetext}
At this stage, we have obtained a perturbative expansion of the first-passage time density for \emph{a particular realisation of} $y$, and only in terms of the expansion coefficients of transition and return probability $T$ and $R$. We now give a more explicit expression of these expansion coefficients.

\subsection{Finding the coefficient terms for probability densities in the functional expansion}
\label{subsec:finding_expansion_terms}
In this section we show how the functional expansion of transition and return probability are obtained perturbatively in terms of some suitable eigenfunctions.

The transition probability $T_{x_0,x_1}(t_0,t_1)$ of the undriven process, characterised by Langevin equation \eqref{eq:driven_langevin} for $\varepsilon=0$, depends on the time-difference only and can therefore be written $T_{x_0,x_1}(t_0,t_1) = T^{(0)}_{x_0,x_1}(t_1-t_0)$; following definitions given in Eqs. \eqref{eq:definition_T_0} and \eqref{eq:def_t_expansion}. The transition density $T^{(0)}$ solves the Kolmogorov forward equation
\begin{align}
\label{eq:Kolmogorov_forward_equation_epsilon_0}
	\begin{cases}
		\partial_t T^{(0)}_{x_0,x_1}(t) &= \LL_{x_1}  T^{(0)}_{x_0,x_1}(t)  \qquad t > 0\\
		T^{(0)}_{x_0,x_1}(t=0) &= \delta(x_1 - x_0) \\
		T^{(0)}_{x_0,x_1}(t)&=0  \qquad t < 0
\end{cases}
\end{align}
where we introduce the forward evolution operator $\LL_x$ as
\begin{align}
	\LL_x \testfnc &= \partial_x (V'(x) \testfnc) + D_x \partial_x^2 \testfnc ,
	\label{eq:forward_equation}
\end{align}
where $\testfnc$ is a twice differentiable test function. In \Eqref{eq:Kolmogorov_forward_equation_epsilon_0}, we denote the forward operator as $\LL_{x_1}$ to indicate that its gradient terms are acting on the $x_1$ dependency of $T^{(0)}$. Correspondingly, the $L^2$-adjoint operator $\LL_x\adj$, also referred to as backward operator, is
\begin{align}
	\LL_x\adj \testfnc &= -V'(x) \partial_x \testfnc + D_x \partial_x^2 \testfnc.
\end{align}
The forward operator $\LL_x$ has a countable set of eigenfunctions $\left\{ u_n(x) \right\}$ and a non-positive spectrum $0 \geq -\lambda_0 > - \lambda_1 > \ldots$ \cite{pavliotis2014},
\begin{align}
	\LL_x u_n(x) = -\lambda_n u_n(x) \qquad n \geq 0
	\label{eq:def_u_n}
\end{align}
but is a priori not self-adjoint in $L^2(\mathbb{R})$ \footnote{Instead, $\LL_x\adj$  is self-adjoint and non-positive in the weighted  space $L^2(u_0(x))$, \cite{pavliotis2014}.}. It is straightforward to show \cite{risken_fokker-planck_1984} however that the  operator
\begin{align}
	\mathfrak{L}_x=e^{ \frac{V(x)}{2 D_x}} \LL_x e^{ - \frac{V(x)}{2 D_x} }
	\label{}
\end{align}
is self-adjoint and that therefore the family of
\begin{align}
	\left \lbrace e^{ \frac{V(x)}{2D_x}} u_n(x) \right \rbrace, n\geq 0
	\label{}
\end{align}
as eigenfunctions of $\mathfrak{L}_x$, form an orthogonal set of eigenfunctions spanning $L^2(\mathbb{R})$. We further impose a choice of normalization of the functions $\{u_n\}$, such that the set $\left \lbrace e^{\frac{V(x)}{2D_x}} u_n(x) \right \rbrace$ is orthonormal.
Defining $u_n(x)$ as right eigenfunctions, and 
\begin{align}
	v_n(x) = e^{\frac{V(x)}{D_x}} u_n(x)
	\label{eq:def_v_n}
\end{align}
as left eigenfunctions, satisfying
\begin{align}
	\LL_x\adj v_n(x) = -\lambda_n v_n(x),
	\label{}
\end{align}
we obtain a biorthogonal system for the pair $\{u_n(x)\}$, $\{v_n(x)\}$:
\begin{align}
	\int_{}^{} \dd x v_m(x) u_n(x) = \delta_{m,n}.
	\label{eq:orthogonality_condition}
\end{align}

This is useful to solve the forward equation; taking the Fourier transform in time of Eq.~\eqref{eq:Kolmogorov_forward_equation_epsilon_0}, one obtains
\begin{align}
	 i\omega \hat{T}^{(0)}_{x_0,x_1}(\omega) = \LL_x\hat{T}^{(0)}_{x_0,x_1}(\omega)+ \delta(x_1 - x_0).
	\label{eq:laplace_trafo_forward}
\end{align}
Inserting the ansatz
\begin{align}
\hat{T}^{(0)}_{x_0,x_1}(\omega)= \sum_{n \geq 0}^{} \hat{T}^{(0)}_{n, x_0}(\omega) u_n(x_1)
	\label{eq:rho_ansatz_zero}
\end{align}
 into \Eqref{eq:laplace_trafo_forward} and using \Eqref{eq:def_u_n} leads to
\begin{align}
	\sum_{n \geq 0}^{}(
i\omega + \lambda_n)\hat{T}^{(0)}_{n, x_0}(\omega) u_n(x_1) = \sum_{n \geq 0}^{} v_n(x_0) u_n(x_1)~,
	\label{eq:ansatz_in_forward_equation}
\end{align}
where we made use of the decomposition of unity,
\begin{align}
	\delta(x_1-x_0) = \sum_{n \geq 0}^{} v_n(x_0)u_n(x_1)~. 
	\label{}
\end{align}
Since the $u_n(x_1)$ are linearly independent, their prefactors in Eq.~\eqref{eq:ansatz_in_forward_equation} need to agree. Therefore,
\begin{align}
	\hat{T}^{(0)}_{n, x_0}(\omega)= \frac{v_n(x_0)}{i\omega+\lambda_n}
	\label{}
\end{align}
implying, together with Eq.~\eqref{eq:rho_ansatz_zero},
\begin{align}
	\hat{T}^{(0)}_{x_0,x_1}(\omega)=\sum_{n \geq 0}^{} \frac{v_n(x_0)u_n(x_1)}{i\omega+\lambda_n}~.
	\label{eq:rho_zero_order_result}
\end{align}

Turning to the case of $\varepsilon\neq 0$, the transition probability of the driven Langevin equation \eqref{eq:driven_langevin} solves the forward equation
\begin{align}
\partial_{t_1} T_{x_0,x_1}(t_0,t_1;[y]) &=\nonumber\\
\left( \LL_{x_1} + \varepsilon y(t_1) \partial_{x_1}\right) &T_{x_0,x_1}(t_0,t_1;[y]), \quad t_1>t_0\nonumber\  \\
T_{x_0,x_1}(t_0, t_0;[y]) &= \delta(x_1 - x_0)\nonumber\\\
T_{x_0,x_1}(t_0, t_1;[y])&=0\quad t_1<t_0
		\label{}
\end{align}
where time-homogeneity can no longer be assumed.
Under Fourier transform, this forward equation becomes
\begin{multline}
	\left( i\omega_1 - \LL_{x_1} \right) \hat{T}_{x_0,x_1}(\omega_0, \omega_1,[\hat{y}])\\
	= \delta(x_1 - x_0)\deltabar(\omega_1+\omega_0)\\
	+  \varepsilon \int_{}^{} \dbar \tilde{\omega}_1 \, \hat{y}(\tilde{\omega}_1) \partial_{x_1}\hat{T}_{x_0,x_1}(\omega_0,\omega_1-\tilde{\omega}_1,[\hat{y}])
	\label{eq:laplace_trafo_forward_y} 
\end{multline}
where the $y$-dependent term turns from a product into a convolution under the Fourier transform.
We develop a perturbative solution of $\hat{T}_{x_0,x_1}(\omega_0,\omega_1,[\hat{y}])$ in powers of $\hat{y}$. 

We first discuss the first-order correction in $\eps$. Following the functional expansion \eqref{eq:functional_expansion_second_exposition}, and using the zeroth order result \eqref{eq:rho_zero_order_result}, we obtain:
\begin{multline}
	\hat{T}_{x_0,x_1}(\omega_0, \omega_1,[\hat{y}]) = \sum_{n \geq 0}^{}\frac{v_n(x_0)u_n(x_1)\deltabar(\omega_0 + \omega_1)}{i\omega_1+\lambda_n} \\
	+  \eps \int_{}^{} \dbar \tilde{\omega}_1 \hat{T}^{(1)}_{x_0,x_1}(\omega_1, \tilde{\omega}_1) \hat{y}( -\tilde{\omega}_1)\deltabar(\omega_0 + \omega_1+\ts_1) + ...
	\label{eq:first_order_expansion_T}
\end{multline}
and we wish to determine the function $\hat{T}^{(1)}$.
Since the $u_n(x)$ span the $L^2$-space, and in analogy to the ansatz \eqref{eq:rho_ansatz_zero},  the first-order correction too can be written as a sum
\begin{align}
    \hat{T}^{(1)}_{x_0,x_1}(\omega_1, \tilde{\omega}_1)= \sum_{n\geq 0}^{} \hat{T}^{(1)}_{n,x_0}(\omega_1,\ts_1) u_n(x_1).
    \label{eq:rho_ansatz_one}
\end{align}
Re-inserting Eqs. \eqref{eq:first_order_expansion_T} and \eqref{eq:rho_ansatz_one} into \Eqref{eq:laplace_trafo_forward_y} causes all terms to zeroth order in $y$ to cancel, and one obtains an equation relating the contributions proportional to $\eps$,
\begin{multline}
	\sum_{n \geq 0}^{} (i\omega_1+ \lambda_n) \int \dbar{\ts_1} \hat{T}^{(1)}_{n, x_0}(\omega_1,\ts_1) \hat{y}(-\ts_1) u_n(x_1) 
	\\
	= \int\dbar{\ts_1} \sum_{n\geq 0}^{} \frac{v_n(x_0) \partial_{x_1}u_n(x_1)}{i(\omega_1 - \ts_1) + \lambda_n} \hat{y} (\tilde{\omega}_1).
	\label{eq:sum_first_order}
\end{multline}
The right hand side, which is the convolution of $\transcoeff{0}{\omega}$ and $y(\omega)$, no longer sums over $u_n(x_1)$ but their derivative $\partial_{x_1}u_n(x_1)$. In order to compare both left and right terms, we need to express this sum as a sum over the linearly independent $u_n(x_1)$ again. The decomposition of the derivative in terms of $u_n(x_1)$ is given by
\begin{align}
	\partial_{x_1} u_n(x_1) = \sum_{k}^{}\deriv_{nk} u_k(x_1)
	\label{eq:gradient_decomposition}
\end{align}
where we refer to the $\deriv_{nk}$ as derivative coupling matrix whose entries, as follows from bi-orthogonality, are
\begin{align}
	\deriv_{nk} = \int_{}^{} \dd x v_k(x) \partial_x u_n(x).
	\label{eq:derivative_coupling_matrix}
\end{align}
Further, \Eqref{eq:sum_first_order} holds true for arbitrary $y(\ts_1)$. In order to compare both integrations over $\dbar \ts_1$, we relabel $\ts_1 \mapsto - \ts_1$ in the left-hand side of the equation.
Using this notation, inserting the sum \eqref{eq:gradient_decomposition} into Eq.~\eqref{eq:sum_first_order}, and resolving the ansatz \eqref{eq:rho_ansatz_one}, one obtains:
\begin{align}
	\hat{T}^{(1)}_{x_0,x_1}(\omega_1,\tilde{\omega}_1) = \sum_{n,k \geq0}^{} \frac{v_k(x_0)\deriv_{kn} u_{n}(x_1) }{(i(\omega_1+\tilde{\omega}_1)+\lambda_k)(i\omega_1 + \lambda_n)}~.
	\label{eq:rho_first_order_result}
\end{align}
In a similar way, the second order correction can be found; using the functional expansion \eqref{eq:functional_expansion_second_exposition} to second order:
\begin{multline}
	\hat{T}_{x_0,x_1}(\omega_0,\omega_1,[\hat{y}]) = \sum_{n \geq 0}^{}\frac{v_n(x_0)u_n(x_1)}{\omega_1+\lambda_n}\deltabar(\omega_0 + \omega_1) \\
	+ \eps \sum_{n,k \geq 0}^{} \frac{v_k(x_0)\deriv_{kn} u_{n}(x_1) }{(-
	i\omega_0+\lambda_k)(i\omega_1 + \lambda_n)} \hat{y}(\omega_0 + \omega_1) \\
	+ \frac{\varepsilon^2 }{2}\iint_{}^{} \dbar{\ts_1} \dbar{\ts_2} \hat{T}^{(2)}_{x_0,x_1}(\omega_1,\ts_1,\ts_2)\hat{y}(-\ts_1)\hat{y}(-\ts_2) \\\deltabar(\omega_0+\omega_1+\ts_1+\ts_2)
	+ \cdots~,
	\label{}
\end{multline}
with the results from Eqs.~\eqref{eq:rho_zero_order_result} and \eqref{eq:rho_first_order_result} to zeroth and first order, inserting this ansatz into the forward equation \eqref{eq:laplace_trafo_forward_y} gives, following in complete analogy to the previous steps,
\begin{multline}
	\hat{T}^{(2)}_{x_0,x_1}(\omega_1, \ts_1, \ts_2) = \\
	\sum_{n,k,m \geq0}^{} \frac{2v_n(x_0)\deriv_{nk}\deriv_{km}u_m(x_1)}{(i(\omega_1\!+\!\ts_1\!+\!\ts_2)+\lambda_n)(i(\omega_1\!+\!\ts_1)+\!\lambda_k)(i\omega_1\!+\!\lambda_{m})}.
	\label{eq:rho_second_order_result}
\end{multline}
Following this method, it is straightforward to generate the perturbative terms of $\hat{T}^{(n)}$  to arbitrary order in $n$,
\begin{multline}
   \hat{T}^{(n)}_{x_0,x_1}(\omega_1,\ts_1,...,\ts_n)
    \\ =n! \sum_{k_0,...,k_n\geq0} \frac{v_{k_0}(x_0)\deriv_{k_0k_1}\cdot...\cdot \deriv_{k_{n-1}k_n}u_{k_n}(x_1)}{ \prod_{j=0}^n \left(i(\omega_1 +\sum_{1\leq \ell\leq n-j} \tilde{\omega}_{\ell}) + \lambda_{k_{j}}\right)}.
    \label{eq:rho_n_order_result}
\end{multline}
Finally, choosing $x_0=x_1$ in any of the expressions \eqref{eq:rho_zero_order_result}, \eqref{eq:rho_first_order_result}, \eqref{eq:rho_second_order_result} and \eqref{eq:rho_n_order_result} gives the corresponding terms for the return probability coefficients $\returncoeff{0}{\omega}, \returncoeff{1}{\omega,\ts_1},...$. Equipped with these expressions, we are able to compute the relevant integrals in the formula for the $y$-averaged first-passage time density \Eqref{eq:averaged_mfpt_general}.

\subsection{Driving noise averaging}
As was set out initially, the quantity of interest is the first-passage time density when averaged over all driving noises (even if the quantity given above might be of interest in itself). The average over driving noise realisations $y_t$ is an average different to the average over the stochastic process $x_t$. 
This double average is reminiscent of ``quenched disorder averages'' where a system subject to thermal fluctuations is embedded into a random disorder potential.  
The expansion in Eq.~\eqref{eq:fpt_expansion_in_y} is a power series in orders of $\eps$, where contributions of order $\eps^n$ contain an internal integration over $n-1$ free frequencies. The expansion terms which stand in front of the $y$ terms, those denoted within square brackets, are independent of $y$. They may be interpreted as the $n$th order response functionals of the first-passage time distribution to perturbations in the driving noise $y$.
To calculate the $y$-average of $\yfpt{\omega_0,\omega_1}$, each term in \Eqref{eq:fpt_expansion_in_y} is integrated over the path-measure of $y$, $\cP[y]$. The order of internal integration and $y$-averaging can be swapped. Consequently, since $\yavg{y_t} = 0$ by assumption, all terms in first order in $y$ vanish. To second order, correlations of $y$ come into play. We introduce the correlation function
\begin{align}
\label{eq:def_correlator}
 \yavg{\hat{y}(\omega_0) \hat{y}(\omega_1)} =	\int \cD[\hat{y}] \cP[\hat{y}] \hat{y}(\omega_0) \hat{y}(\omega_1)=\corr{\omega_1} \deltabar(\omega_0 + \omega_1) ,
\end{align}
where the last equality arises from the stationary in time of the random process $y(t)$. Stationarity in time also implies that $\corr{\omega}$ is symmetric in $\omega \mapsto - \omega$. Performing an averaging over $\cP[\hat{y}]$ of \Eqref{eq:fpt_expansion_in_y}, we obtain:
\begin{widetext}
\begin{multline}
\label{eq:averaged_mfpt_general}
    \yavg{\hat{p}(\omega_0,\omega_1;[\hat{y}])} = \frac{\transcoeff{0}{\omega_1}}{\returncoeff{0}{\omega_1}}\deltabar(\omega_0+\omega_1) \\
    +\frac{1}{2} \eps^2 \left[-\underbrace{\int \dbar{\ts} \frac{ \transcoeff{0}{\omega_1}\returncoeff{2}{\omega_1,\ts,-\ts}}{\left(\returncoeff{0}{\omega_1}\right)^2}\corr{\ts} }_{=:({\rm I})}  
    + \underbrace{ 2 \int \dbar{\ts} \frac{\transcoeff{0}{\omega_1}\returncoeff{1}{\omega_1-\ts,\ts}\returncoeff{1}{\omega_1,-\ts}}{\left(\returncoeff{0}{\omega_1}\right)^2 \returncoeff{0}{\omega_1-\ts}} \corr{\ts} }_{=:({\rm II})} 
    \right. \\
    \left.
    - \underbrace{2\int \dbar{\ts} \frac{\transcoeff{1}{\omega_1-\ts,\ts}\returncoeff{1}{\omega_1,-\ts}}{\returncoeff{0}{\omega_1}\returncoeff{0}{\omega_1-\ts}}\corr{\ts}}_{=:(\rm{III)}} + \underbrace{\int \dbar{\ts}\frac{\transcoeff{2}{\omega_1,\ts,-\ts}}{\returncoeff{0}{\omega_1}}\corr{\ts}}_{=:({\rm IV})}\right]\deltabar(\omega_0+\omega_1) + ...
\end{multline}
\end{widetext}
The first term, of zeroth order, represents the Darling-Siegert solution \eqref{eq:FPTMGFclassic}. This is consistent with our expansion around the  base-point of no driving noise (fully Markovian process). Once averaged, the second-order contribution is again diagonal (\ie proportional to $\deltabar(\omega_0 + \omega_2)$) indicating that the $y$-averaged first-passage distribution is again invariant under time-shifts. The four correction terms featuring in the second order expansion in $\eps$ are labelled $\rm (I)$ to $\rm (IV)$, and need to be  calculated explicitly. 

The corresponding expressions are derived in the following section, for the case of Gaussian coloured noise.
Generally, however, the entire expression \Eqref{eq:averaged_mfpt_general} remains valid for driving noise correlations $\corr{\omega}$ which keep the integrals finite.

\subsection{Second-order correction with Gaussian coloured noise}
So far, in our derivation of the second-order correction to the first-passage time density, \Eqref{eq:averaged_mfpt_general}, we only demanded the active driving noise to be stationary, with finite correlations and vanishing mean. In what follows, we specify $y_t$ to be Gaussian coloured noise. This choice is almost canonical  in the study of coloured noise \cite{hanggi_colored_1995}. In our case it greatly simplifies the necessary integrals. It is, however, possible to use any other correlation functions as long as the integrals remain manageable. Generally, to compute the perturbative contribution of $n$th order in $y$, the $n$-point correlation function of $y_t$ needs to be known; For Gaussian processes all higher moments follow from the two-point correlation function which simplifies the calculation of potential higher order corrections. Since the correlation function of coloured noise is an exponential, the results obtained in this section to order $\eps^2$ hold for \emph{any} noise with such autocorrelation, in particular telegraphic noise as used in run-and-tumble processes \cite{cates_diffusive_2012}. Run-and-tumble and other purely active processes correspond to $D_x = 0$, or alternatively $\nu \to \infty$, and are therefore not captured by our perturbative approach valid for $\nu \lesssim 1$.

Gaussian coloured noise is defined by its exponential correlator,
\begin{align}
    \yavg{y_{t_0} y_{t_1}} = D_y \beta e^{-\beta|t_1 - t_0|},    \label{eq:colerator} 
\end{align}
which in Fourier space reads (\Eqref{eq:def_correlator})
\begin{align}
    \corr{\omega} = \frac{2D_y \beta^2}{\omega^2+\beta^2}~.
\end{align}
With the explicit expressions \eqref{eq:rho_zero_order_result},\eqref{eq:rho_first_order_result}, \eqref{eq:rho_second_order_result}, we perform the integration in ${\rm (I)}$ (see \Eqref{eq:averaged_mfpt_general} for notation) in eigenfunction-representation,
\begin{align}
\label{eq:diagram_I}
    {\rm (I)} &= \int \dbar{\ts} \frac{ \transcoeff{0}{\omega_1}\returncoeff{2}{\omega_1,\ts,-\ts}}{\left(\returncoeff{0}{\omega_1}\right)^2}\corr{\ts} \nonumber \\
    & = 2D_y \beta^2 \,  \frac{ \transcoeff{0}{\omega_1}}{\left(\returncoeff{0}{\omega_1}\right)^2}\int \dbar{\ts} \frac{\returncoeff{2}{\omega_1,\ts,-\ts}}{\beta^2+\ts^2} \nonumber \\
    & = 2D_y \beta^2\,  \frac{ \transcoeff{0}{\omega_1}}{\left(\returncoeff{0}{\omega_1}\right)^2}  \sum_{n,k,m \geq 0} \nonumber
    \\
    &\cdot2 \int_{-\infty}^{\infty} \dbar{\ts} \frac{v_n(x_1)\deriv_{nk}\deriv_{km}u_m(x_1)}{(i\omega_1 + \lambda_n)(i(\omega_1 + \ts) + \lambda_k)(i\omega_1 + \lambda_m)}\frac{1}{\ts^2+\beta^2} \nonumber \\
    & = D_y  \beta  \frac{ \transcoeff{0}{\omega_1}\returncoeff{2}{\omega_1,-i\beta,i\beta}}{\left(\returncoeff{0}{\omega_1}\right)^2} .
\end{align}
where in the last equality we employed Cauchy's residue theorem closing the contour in the lower half-plane containing the simple pole at $\tilde{\omega} = -i\beta$. Likewise, we find
\begin{align}
    {\rm (IV)} &= \label{eq:diagram_IV} \int \dbar{\ts}\frac{\transcoeff{2}{\omega_1,\ts,-\ts}}{\returncoeff{0}{\omega_1}}\corr{\ts} \nonumber\\
    & = 2 D_y \beta^2  \frac{1}{\returncoeff{0}{\omega_1}}  \sum_{n,k,m \geq 0} \nonumber \\
    & \cdot 2\int_{\infty}^{\infty} \dbar{\ts} \frac{v_n(x_0)\deriv_{nk}\deriv_{km}u_m(x_1)}{(i\omega_1 + \lambda_n)(i(\omega_1 + \ts) + \lambda_k)(i\omega_1 + \lambda_m)}\frac{1}{\ts^2+\beta^2} \nonumber \\
    & = D_y \beta  \frac{\transcoeff{2}{\omega_1,-i\beta,i\beta}}{\returncoeff{0}{\omega_1}}
\end{align}
where again the integral is evaluated by closing the contour in the lower half-plane enclosing the pole at $\tilde{\omega}=-i\beta$.

The integrals ${\rm (II)}$ and ${\rm (III)}$, featuring $\ts$-dependent denominators, require some more careful analysis.
We have
\begin{align}
	{\rm (II)} &= 2 \int \dbar{\ts} \frac{\transcoeff{0}{\omega_1}\returncoeff{1}{\omega_1-\ts,\ts}\returncoeff{1}{\omega_1,-\ts}\corr{\ts} }{\left(\returncoeff{0}{\omega_1}\right)^2 \returncoeff{0}{\omega_1-\ts}} ~.
\end{align}
As before we calculate this integral by application of Cauchy's residue theorem, but now closing the contour in the upper half-plane. The numerator's poles all lie in the lower half-plane with the exception of the pole at $\tilde{\omega}=i\beta$ stemming from the correlator, as can be confirmed by inspection of \Eqref{eq:rho_first_order_result}. Besides, the denominator $\returncoeff{0}{\omega_1 - \tilde{\omega}}$ does not have any roots for $\Im(\tilde{\omega}) >0$. Indeed, using the relations \Eqref{eq:rho_zero_order_result} and \eqref{eq:def_v_n}, $\hat{R}^{(0)}(\omega_1-\ts)=\sum_{n \geq 0}^{} \frac{v_n(x_1)u_n(x_1)}{i(\omega_1-\ts)+\lambda_n}= e^{-V(x_1)/D} \sum_{n \geq 0}^{} \frac{\left(u_n(x_1)\right)^2}{i(\omega_1-\ts)+\lambda_n}$. Since the $\{u_n(x)\}$ span $L^2$, there cannot be a $x_1$ for which all $u_n(x_1) = 0$. Besides the real part of the denominator in the sum is strictly positive, $\lambda_n +  \Im(\tilde{\omega})>0$, assuming $\omega_1 \in \mathbb{R}$. In the upper half plane the sum therefore only contains terms whose real part is non-negative and at least  once strictly positive; hence the sum  is free of roots in the upper half plane. Invoking Cauchy's residue formula the integral is then given by
\begin{align}
 {\rm    (II) }=2D_y \beta  \frac{\transcoeff{0}{\omega_1}\returncoeff{1}{\omega_1-i\beta,i\beta}\returncoeff{1}{\omega_1,-i\beta}}{\left(\returncoeff{0}{\omega_1}\right)^2 \returncoeff{0}{\omega_1-i\beta}}\ .
    	\label{eq:diagram_II}
\end{align}
By analogous reasoning one obtains
\begin{align}
{\rm	(III) }&= 2\int \dbar{\ts} \frac{\transcoeff{1}{\omega_1-\ts,\ts}\returncoeff{1}{\omega_1,-\ts}}{\returncoeff{0}{\omega_1}\returncoeff{0}{\omega_1-\ts}}\corr{\ts}\nonumber\\
	&=2D_y\beta \frac{\transcoeff{1}{\omega_1-i\beta,i\beta}\returncoeff{1}{\omega_1,-i\beta}}{\returncoeff{0}{\omega_1}\returncoeff{0}{\omega_1-i\beta}}~.
	\label{eq:diagram_III}
\end{align}
All four terms together give a general formula for the moment generating function of first-passage times for \emph{arbitrary underlying processes and driving noises} provided the eigenfunctions and correlators are known. In the case of driving noise with exponentially decaying auto-correlation, the full formula for $\yavg{\fpt{\omega_0,\omega_1}} = \hat{F}({\omega_1})\deltabar(\omega_0+\omega_1)$, reads
\begin{widetext}
\begin{align}
	\label{eq:fpt_correction_general}
	\hat{p}({\omega}) &= \frac{
		\transcoeff{0}{\omega}}
		{
	\returncoeff{0}{\omega}
	}
	+ \frac{\eps^2 D_y  \beta }{2 \returncoeff{0}{\omega}}
	\left[  \frac{ \transcoeff{0}{\omega}}{\returncoeff{0}{\omega}} \left(2\frac{\returncoeff{1}{\omega-i\beta,i\beta}\returncoeff{1}{\omega,-i\beta}}{ \returncoeff{0}{\omega-i\beta}} -\returncoeff{2}{\omega,-i\beta,i\beta}\right)\right.\nonumber \\
	&\left.\qquad \qquad  \qquad+\transcoeff{2}{\omega,-i\beta,i\beta} -2\frac{\transcoeff{1}{\omega-i\beta,i\beta}\returncoeff{1}{\omega,-i\beta}}{\returncoeff{0}{\omega-i\beta}} 
	\right]+ \cO(\eps^4)~.
\end{align}
\end{widetext}
This general result concludes this section. In the next section, we consider two concrete examples to demonstrate how this perturbation theory can be turned into analytical results.

\section{\label{sec:results} Results for simple potentials}

\subsection{Active Thermal Ornstein Uhlenbeck Process (ATOU)}
\label{subsec:ouou}
In this example we study the case of a  particle in a harmonic potential driven by white and coloured noise described by the Langevin Equation
\begin{align}
	\dot x_t = - \alpha x_t + \xi_t + \varepsilon y_t
	\label{eq:ouou_SDE}
\end{align}
with driving noise correlator (see \Eqref{eq:colerator})
\begin{align}
	\yavg{y_{t_0} y_{t_1}} = D_y \beta e^{-\beta|t_1-t_0|}.
\end{align}
This process reduces to the simple Ornstein Uhlenbeck process when $\eps = 0$ which models a particle in a harmonic potential $(V(x)=\frac{\alpha}{2}x^2$) within a thermal bath. We consider, however, the process driven by an additional ``active'' term $\eps y_t$. We therefore refer to this process as active thermal Ornstein Uhlenbeck process (ATOU). 
In the undriven case ($\eps=0$), the dynamics are characterised by the time and length-scales $\alpha^{-1}$ and $$\ell = \sqrt{D_x \alpha^{-1}}.$$

\subsubsection{From eigenfunctions to the moment generating function of first-passage times\label{subsec:OUOU_eigenfunctions}}
The Fokker-Planck equation associated (\cf \Eqref{eq:forward_equation}) to the Langevin Equation  \eqref{eq:ouou_SDE} 
is
\begin{align}
\partial_t T^{(0)}_{x_0,x_1} = 
\mathcal{L}_{x_1}T^{(0)}_{x_0,x_1} &= \left( D_x\partial_{x_1}^2+ \alpha x_1 \partial_{x_1} + \alpha  \right)  T^{(0)}_{x_0,x_1}~.
\end{align}
with $\mathcal{L}_x =  D_x\partial_{x}^2+ \alpha x \partial_{x} + \alpha  $ the forward evolution operator.
The operator has eigenvalues \cite{pavliotis2014}
\begin{align}
	\lambda_n = \alpha n
\end{align}
and is diagonalised by the (normalised) eigenfunctions
\begin{align}
	v_n(x) & = \frac{1}{\sqrt{\sqrt{2 \pi}\ell \cdot n!}} \He_n\left( \frac{x}{\ell} \right)\\
	u_n(x) & = \frac{1}{\sqrt{\sqrt{2 \pi}\ell \cdot n!}} \He_n\left( \frac{x}{\ell} \right)\cdot e^{ -\frac{x^2}{2 \ell^2}}
	\label{}
\end{align}
where we introduced Hermite polynomials using the convention
\begin{align}
	\He_n(x) = (-1)^n e^{\frac{{x}^2}{2}} \frac{\mathrm{d}^n}{\mathrm{d} x^n} e^{-\frac{{x}^2}{2}} 
	\label{}
\end{align}
which satisfy the relation
\begin{align}
	\partial_{x}\He_n(x) = x\He_n(x) - \He_{n+1}(x)
	\label{}
\end{align}
such that the coupling matrix (cf.~\Eqref{eq:derivative_coupling_matrix}) resolves to
\begin{align}
	\deriv_{mn} = -\frac{\sqrt{n}}{ \ell}\delta_{m+1, n}~.
	\label{eq:ou_derivative_coupling}
\end{align}
The coupling matrix is not diagonal: incoming momentum $n$ is upgraded to outgoing momentum $n+1$ by the noise coupling. The amplitude of $\deriv_{m,n}$ grows like $|\deriv_{m,n}| \sim \sqrt{\lambda_m }$. For later use, we also note that
\begin{align}
	\left(\frac{x}{\ell} -\ell  \partial_x\right) v_n(x) &= \sqrt{n+1} \, v_{n+1}(x)  \label{eq:v_creation}\\
	\ell \partial_x v_n(x) &= \sqrt{n} \, v_{n-1}(x)~.
	\label{eq:v_annihilation}
\end{align}
By $L^2$-adjointness, it follows that the adjoint creation and annihilation operators are
\begin{align}
	-\ell \partial_x u_{n}(x) &= \sqrt{n+1} \,u_{n+1}(x)  \label{eq:u_creation}\\
	\left(\frac{x}{\ell}+ \ell\partial_x\right) u_n(x) &= \sqrt{n} u_{n-1}(x) ~.
	\label{eq:u_annihilation}
\end{align}

In order to compute the transition and return probabilities, the following identity \cite{MagnusOberhettingerSoni:1966}
\begin{multline}
	\sum_{k = 0 }^{\infty} \frac{\He_k(x) \He_k(y)e^{-\frac{y^2}{2}}}{k!} z^k = \frac{1}{ \sqrt{ 1 - z^2}} e^{ -\frac{(y-zx)^2}{2(1-z^2)}}
	\label{eq:hermite_identity}
\end{multline}
proves to be useful.

We introduce all quantities in dimensionless form, and take $\omega$ to be imaginary, such as the reduced frequency $\bar{s}$ and the reduced autocorrelation time $\bar\beta$ as
\begin{align}
	\bar{s} = i \alpha^{-1} \omega \qquad \bar\beta = \alpha^{-1} \beta
	\label{}
\end{align}
By considering the Fourier transform at $\bar{s} = i \alpha^{-1}\omega$, we are effectively studying the Laplace transform. This is intended since our final observable is the moment generating function, the Laplace transform of the first-passage time distribution. 
Further, we denoted the lengths rescaled by $\ell$ as
\begin{align}
\target = \ell^{-1}x_1 \qquad \start = \ell^{-1}x_0.
\end{align}
In the discussion that follows, we analyse all densities as densities in these dimensionless quantities to simplify notation and calculations. 
Following \Eqref{eq:rho_zero_order_result}, one obtains for the transition probability
\begin{align}
	&\transcoeff{0}{\omega=-i\bar{s}\alpha} \\
	&= \frac{1}{\sqrt{2 \pi}\ell} \sum_{n=0}^{} \frac{\He_n(\start) \He_n(\target) e^{-\frac{ {\target}^2}{2}}}{n! (\bar{s} \alpha+\alpha n)}  \nonumber \\
	& = \frac{1}{\sqrt{2 \pi} \ell \alpha} \int_{0}^{\infty} \dd{t}\, e^{-\bar{s} t} \sum_{n=0}^{\infty} \frac{ \He_n(\start)\He_n(\target) e^{-\frac{ {\target}^2}{2}}}{n!}\left( e^{-t} \right)^n \nonumber \\
	& = \frac{1}{\sqrt{2 \pi}\ell \alpha} \int_{0}^{\infty} \dd{t} \frac{e^{-\bar{s} t }}{\sqrt{1-e^{-2t}}}e^{ -    \frac{(\target - \start e^{-t})^{2}}{2(1 - e^{-2t})} } \nonumber 
	\label{}
\end{align}
where we used identity \eqref{eq:hermite_identity} setting $z = e^{-t}$. This integral is the Laplace transform of the Ornstein-Uhlenbeck propagator (in $t \leftrightarrow s$), and its value is known in the literature to be (\cite{siegert_first_1951})
\begin{align}
	&\transcoeff{0}{-i\bar{s}\alpha} \label{eq:ouou_rho0xy} \\
	&= \begin{cases}
		\frac{\Gamma(\bar{s})}{\sqrt{2 \pi }\ell \alpha }e^{ \frac{ {\start}^2 - {\target}^2}{4}} D_{-\bar{s}}\left( -\start \right) D_{-\bar{s}}\left( \target\right)  & \start < \target \\
	\frac{\Gamma(\bar{s})}{\sqrt{2 \pi} \ell  \alpha } e^{\frac{ {\start}^2 - {\target}^2}{4}}D_{-\bar{s}}\left( \start \right) D_{-\bar{s}}\left( - \target \right)  & \start > \target
	\end{cases} \nonumber
\end{align}
where we introduced the parabolic cylinder functions $D_{-\bar{s}}(x)$ (\cite{gradshteyn_table_2007}).  By continuity, for $\start \to \target$ it follows that
\begin{align}
	\returncoeff{0}{-i\bar{s}\alpha} = \frac{\Gamma(\bar{s})}{\sqrt{ 2 \pi } \ell \alpha} D_{-\bar{s}}(\target)D_{-\bar{s}}(-\target).
	\label{eq:ouou_rho0yy}
\end{align}
In order to compute $\transcoeff{1}{-i\bar{s} \alpha, -i\bar{\beta}\alpha}$, we use Eq.~\eqref{eq:rho_first_order_result} and the derivative coupling matrix computed in \eqref{eq:ou_derivative_coupling} to find
\begin{align}
	\label{eq:ouou_rho1xy}
&\transcoeff{1}{-i\bar{s}\alpha,- i\bar{\beta}\alpha} \\
&= \frac{1}{ \ell \alpha^2} \sum_{n}^{} \frac{v_{n}(\start) (-\sqrt{n+1})u_{n+1}(\target) }{(\bar{s} + \bar{\beta}  + n )(\bar{s} + n + 1)}\nonumber\\
&	= \frac{1}{\ell \alpha^2(1-\bar{\beta})} \nonumber\\
	&\qquad \times \partial_{\target} \sum_{n}^{} v_{n}(\start)u_n(\target) \left[\frac{1}{\bar{s}  + \bar{\beta}  +n} -\frac{1}{\bar{s} +n+1 } \right] \nonumber\\
	&= \frac{\partial_{\target}}{\ell  \alpha  (  1- \bar{\beta})} \left[ \transcoeff{0}{-i(\bar{s} + \bar{\beta})\alpha} - \transcoeff{0}{-i(\bar{s}  + 1)\alpha}   \right] \nonumber
\end{align}
where we made use of relation \eqref{eq:u_creation} in the second equality. Letting $\start \to \target$, one obtains $\returncoeff{1}{ -i\bar{s}\alpha, -i\bar{\beta}\alpha} $. The counterpart, $\transcoeff{1}{ -i\bar{s}\alpha-i\bar{\beta}\alpha,i\bar{\beta}\alpha}$, is similarly found to be
\begin{multline}
	\transcoeff{1}{ -i\bar{s}\alpha-i\bar{\beta}\alpha,i\bar{\beta}\alpha}\\
	= \frac{\partial_{\target}}{\ell \alpha (1 + \bar{\beta})}\left[ \hat{T}^{(0)}( -i\bar{s}\alpha) - \hat{T}^{(0)}( -i(\bar{s} + \bar{\beta} + 1) \alpha ) \right]
	\label{}
\end{multline}
These terms can be explicitly calculated and simplified. The rather lengthy but explicit expressions are given in appendix \ref{app:explicit_densities_ou}.

For the second order derivative term, using formula~\eqref{eq:rho_second_order_result}, one finds
\begin{align}
	&\transcoeff{2}{ -i\bar{s}\alpha, -i\bar{\beta}\alpha, i\bar{\beta}\alpha} 	\label{eq:ouou_rho2xy}  \\
	&= \frac{2}{ \ell^2  \alpha^3} \sum_{n=0}^{\infty} \frac{v_{n}(\start)\sqrt{n+1}\sqrt{n+2} u_{n+2}(\target)}{(\bar{s}+n)(\bar{s}+\bar{\beta}+n+1)(\bar{s}+n+2)} \nonumber \\
	& = \frac{2\partial_{\target}^2}{ \ell^2 \alpha^3} \sum_{n=0}^{\infty} \frac{v_n(\start)u_n(\target)}{(\bar{s} + n)(\bar{s} + \bar{\beta} + n + 1)(\bar{s} + n + 2)} \nonumber \\
	& = \frac{2\partial_{\target}^2}{\ell^2  \alpha^2} \left[\frac{1}{2(\bar{\beta}+1)}\transcoeff{0}{-i\bar{s}\alpha} - \frac{1}{2(\bar{\beta}-1)}\transcoeff{0}{-i(\bar{s}+2)\alpha} \right. \nonumber \\
	& \qquad \left. + \frac{1}{({\bar{\beta}}^2 -1)} \transcoeff{0}{-i(\bar{s}+\bar{\beta}+1)\alpha} \right] \nonumber
\end{align}
Again, the evaluated terms, including for $\start \to \target$ are given in appendix \ref{app:explicit_densities_ou}.

Equipped with the return and transition probabilities and its first two derivatives with respect to driving noise $y$ (cf.~\eqref{eq:ouou_rho0xy}-\eqref{eq:ouou_rho2xy}), we  obtain the four contributions \eqref{eq:diagram_I}-\eqref{eq:diagram_III} which constitute the second-order correction formula \eqref{eq:fpt_correction_general}. Whilst all the explicit expressions are given in appendix \ref{app:explicit_densities_ou}, we here give the moment generating function in full as a undriven part and a perturbative correction, using $s = \alpha \bar{s} = i \omega$,
\begin{multline}
M_{x_0,x_1}(s)=M^{0, {\rm OU}}_{\frac{x_0}{\ell}, \frac{x_1}{\ell}}\left(\frac{s}{\alpha}\right) 
+ \underbrace{\frac{D_y \varepsilon^2 \beta}{D_x \alpha}}_{=:\nu}M^{1, {\rm OU}}_{\frac{x_0}{\ell}, \frac{x_1}{\ell}}\left( \frac{s}{\alpha}  \right) + \cO(\nu^2)
\label{eq:atou_mgf}
\end{multline}
where we introduced the dimensionless parameter of expansion $
    \nu = \frac{\eps^2 D_y\beta }{D_x \alpha}.$

As is already known from literature (e.g.~\cite{darling_first_1953}),
\begin{align}
M^{0, {\rm OU}}_{\start, \target} (\bar{s}) = \begin{cases}
 e^{\frac{{\start}^2-{\target}^2}{4} }\frac{D_{-\bar{s}}(-\start)}{D_{-\bar{s}}(-\target)} & \start < \target \\
e^{\frac{{\start}^2-{\target}^2}{4} }\frac{D_{-\bar{s}}(\start)}{D_{-\bar{s}}(\target)} & \start > \target.
 \end{cases}
 \label{eq:OUOU_M0}
\end{align}

By symmetry $(\start, \target) \leftrightarrow (-\target,-\start)$ of the problem and symmetry of driving noise, it suffices to regard one case only, such that without loss of generality we assume $\start < \target$.
The central result of this section then is
\begin{widetext} 
\begin{align}
M^{1, {\rm OU}}_{\start, \target} (\bar{s}) &
=  \frac{\bar{s}  e^{\frac{{\start}^2-{\target}^2}{4}}}{2 \left({\bar{\beta}} ^2-1\right) D_{-\bar{s} }(-\target){}^2 D_{-{\bar{\beta}} -\bar{s} }(-\target)}
\nonumber\\
&\times \left[ \left({\bar{\beta}} +1\right)(\bar{s} + 1)D_{-{\bar{\beta}}-\bar{s}}(-\target) \left(D_{-\bar{s}}(-\start)D_{-\bar{s}-2}(-\target)-D_{-\bar{s}-2}(-\start)D_{-\bar{s}}(-\target) \right)\right. 
\nonumber\\
 & \qquad \left. -2 ({\bar{\beta}} +\bar{s} ) D_{-\bar{s} -1}(-\target) \left(D_{-\bar{s} }(-\start) D_{-{\bar{\beta}} -\bar{s} -1}(-\target)- D_{-{\bar{\beta}} -\bar{s} -1}(-\start)D_{-\bar{s} }(-\target)\right)\right]
\label{eq:OUOU_M1}
\end{align}
\end{widetext}
Using for instance a computer algebra system like \cite{Mathematica}, all moments can be obtained by differentiation and evaluating the limit of $\bar{s} \to 0$ at which all derivatives have a removable singularity.

\subsubsection{Numerical Validation}
\label{subsubsec:numerical_validation_ouou}
\begin{figure*}
 \subfloat[\textbf{Correction to mean first passage time of ATOU} (\cf \Eqref{eq:ouou_SDE}) as obtained from \Eqref{eq:def_cT1_ouou} versus target positions $x_1$, $x_0=0$ fixed, and various values of $\nu$ (\textbf{plot marks}) compared to theoretical result to first order in $\nu$ (\textbf{black line}) using  \Eqref{eq:how_to_t_scaling} and the result obtained in \eqref{eq:OUOU_M1}. The \textbf{inset} shows the mean first-passage time $\fptxy$ as measured vs $x_1$ for values of $\nu = 0$ to $0.8$. Correction due to active driving noise increases MFPT for $x_1 \lesssim 1.6$ and decreases MFPT for $x_1 \gtrsim 1.6$. This behaviour is fully captured by the analytic result.
  ]{\includegraphics[width=1.0\columnwidth]{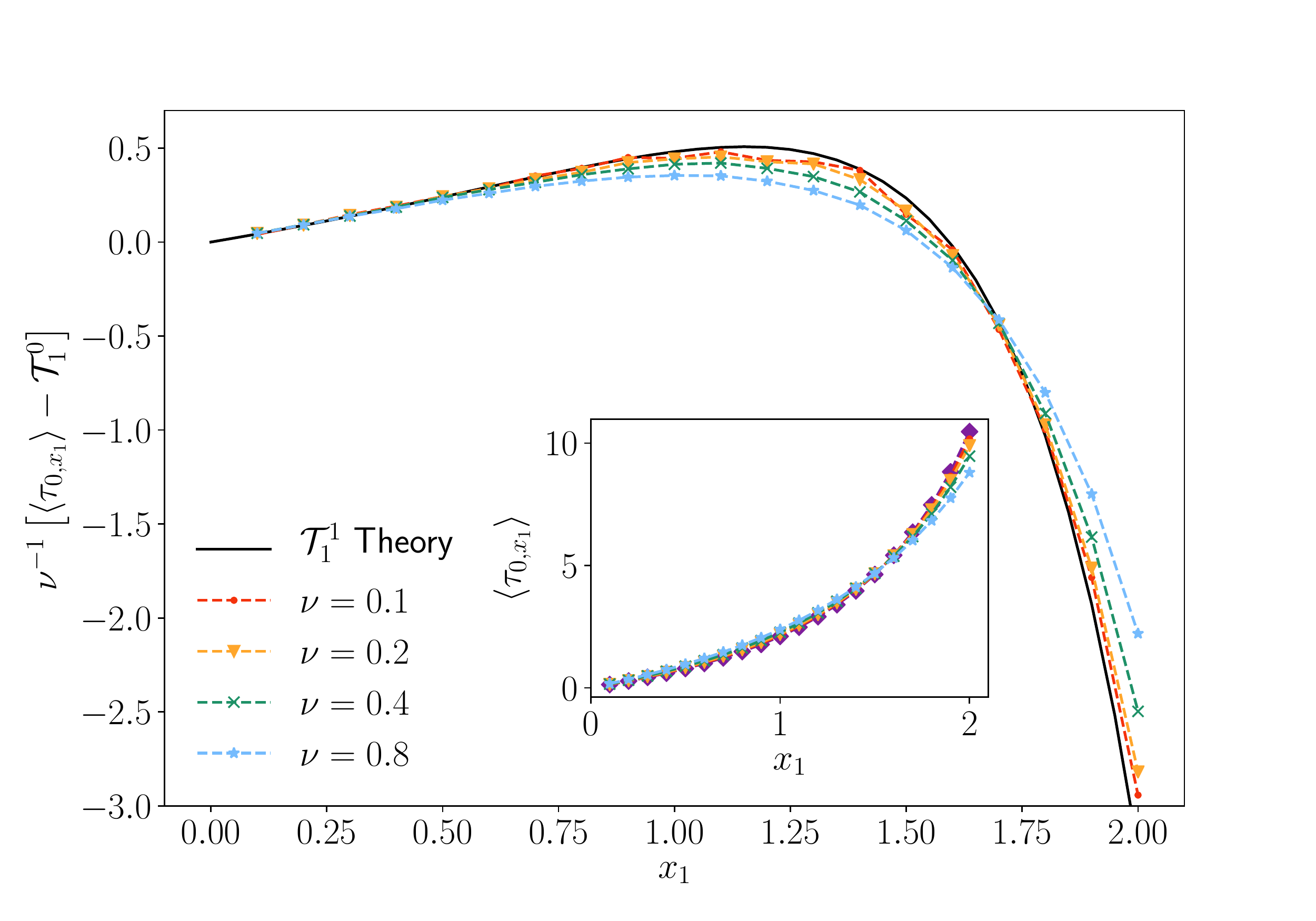}\label{fig:mom1_fpt_ouou}}\hfill
  \subfloat[\textbf{Correction to mean squared first passage time of ATOU} (\cf \Eqref{eq:ouou_SDE}) as obtained from \Eqref{eq:def_cT2_ouou} versus target positions $x_1$, $x_0=0$ fixed, and various values of $\nu$ (\textbf{plot marks}) compared to theoretical result to first order in $\nu$ (\textbf{black line}) using  \Eqref{eq:how_to_t_scaling} and the result obtained in \eqref{eq:OUOU_M1}. The \textbf{inset} shows the mean squared first-passage time $\fptxy$ as measured vs $x_1$ for values of $\nu = 0$ to $0.8$. ]{\includegraphics[width=1.0\columnwidth]{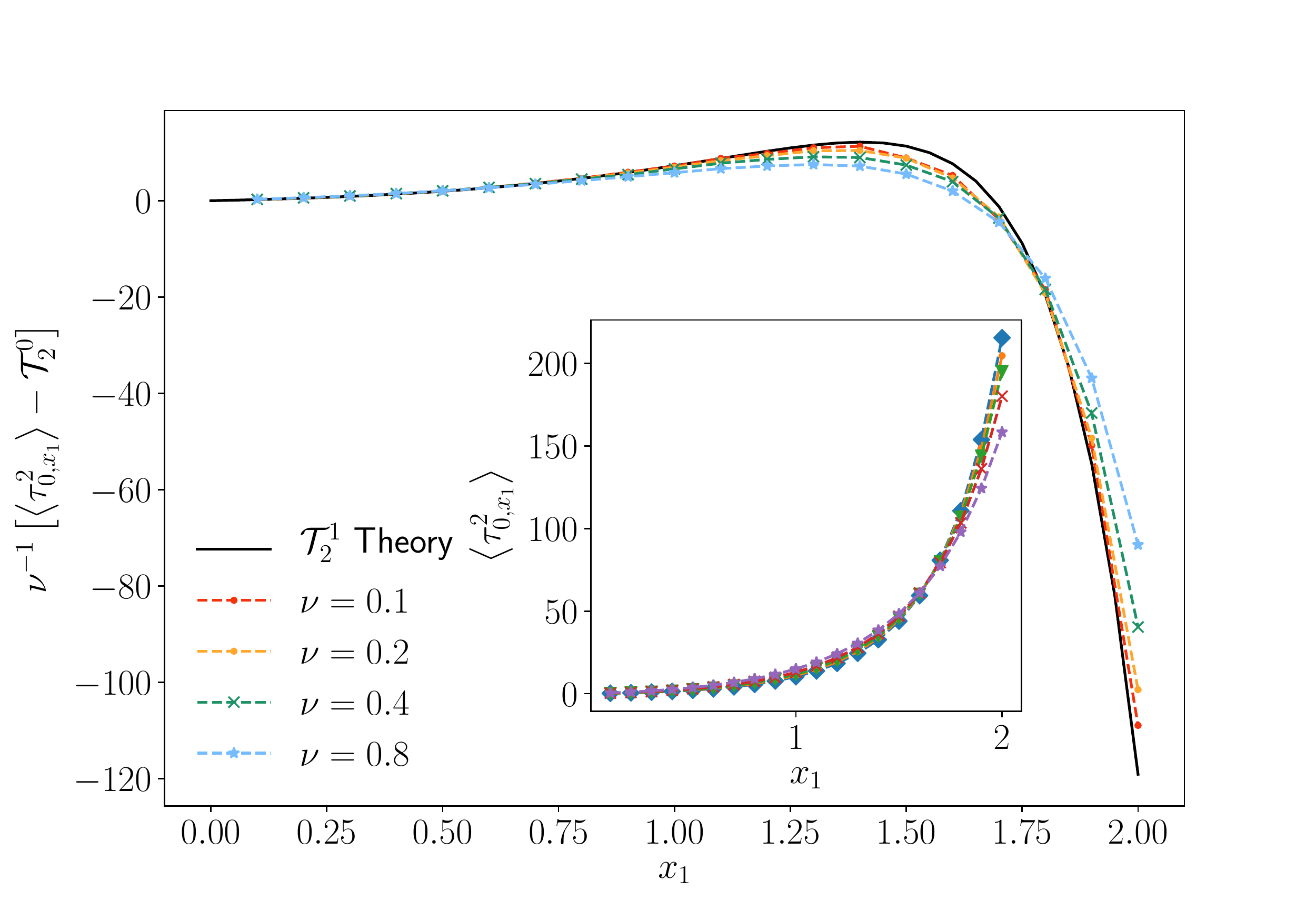}\label{fig:mom2_fpt_ouou}}
  \caption{First order correction to first and second moment of ATOU. Simulation parameters are $x_0 = 0, D_x = 1, D_y = 1, \alpha = 1, \beta = \frac{1}{2}$ and $\eps$ suitably chosen to fix $\nu = D_y \beta \eps^2/(D_x \alpha)$. Averages were taken over $10^6$ samples.}
\end{figure*}
\begin{figure}
\includegraphics[width=1.0\columnwidth]{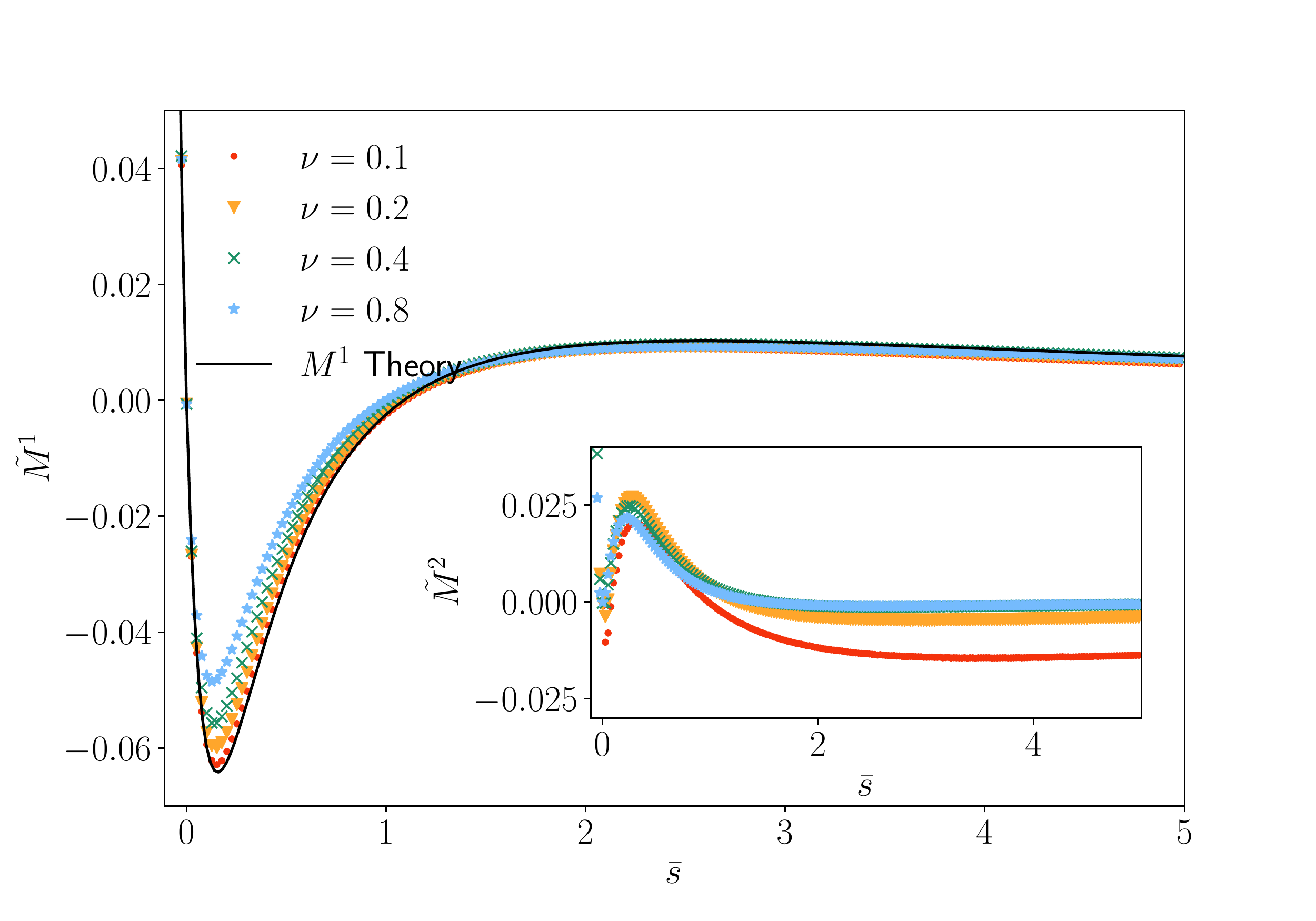}
\caption{\textbf{Numerical validation of first order correction to FPT moment generating function} $M^1$ (\cf \Eqref{eq:general_f_structure}) of ATOU (see \Eqref{eq:ouou_SDE} and \secref{subsec:ouou} for discussion). The result is calculated in \Eqref{eq:OUOU_M1}. Numerical Simulations are shown for various values of $0.1 \leq \nu \leq 0.8$  (\textbf{plot marks}). The moment generating functions were sampled for $x_0=0, x_1 = 1, D_x = 1, D_y = 1, \alpha = 1, \beta = \frac{1}{2}$ and $\eps$ suitably chosen to fix $\nu = D_y \varepsilon^2 \beta /(D_x \alpha)$. For small values of $\nu$ agreement with theoretical first-order correction (\textbf{black line}) is excellent. For larger values of $\nu$ the deviation  increases. The rescaled deviation, $\tilde{M}_2$ (see \Eqref{eq:def_cM2_ouou}), (\textbf{inset}) collapse and thus confirm that these deviations are systematic higher-order corrections. See \eqref{subsubsec:numerical_validation_ouou} for further results and discussion.}
\label{fig:ouou_m1}
\end{figure}

In order to corroborate the closed form result of the first-order correction to the moment generating function of first-passage times, Eq.~\eqref{eq:OUOU_M1}, we employ Monte Carlo simulations integrating the driven Langevin equation \eqref{eq:ouou_SDE} $N \simeq 10^6$ times, numerically find the first-passage time $\tilde{\tau}_i$, and average the moment generating function  $\tilde{M} = \frac{1}{N}\sum_{i=1}^N e^{-s \tilde{\tau}_i}$ for values of $s$ within the range $s\in [0,5]$. In the following, $\tilde{M}$ and $\tilde{\cT}$ denote quantities that have been numerically obtained. Since we assume an expansion of the form $M = M^0 + \nu M^1+ \nu^2 M^2 + ...$, we take the numerical first derivative
\begin{align}
    \tilde{M}_{1}|_{\nu} = \frac{\tilde{M}|_{\nu}-\tilde{M}|_{\nu=0}}{\nu}
\end{align}
to verify our analytic prediction of $M^1$. In Fig.~\ref{fig:ouou_m1}, the numerical estimate $\tilde{M}_1$ is shown for various values of $\nu$, together with the analytic expression \Eqref{eq:OUOU_M1} of the scaling function $M^1$. For small $\nu$, the agreement is excellent. For larger values of $\nu$, higher-order corrections become more visible. The next-higher contribution, which we did not calculate analytically but which can be found by following the framework to second order in $\nu$, is numerically estimated by taking the second numerical derivative,
\begin{align}
    \tilde{M}^2|_{\nu} = \frac{\tilde{M}^1|_{\nu}  -M^1}{\nu},
    \label{eq:def_cM2_ouou}
\end{align}
and is shown in the inset of Fig.~\ref{fig:ouou_m1}. For $0.2 \leq \nu \leq 0.8$, the second-order corrections collapse, indicating that the deviations in the main figure are well accounted for by second-order corrections. For $\nu=0.1$, $\tilde{M}^2$ deviates slightly due to the statistical noise, since the second order correction is very small. 

For better physical intuition of the process, we numerically compute the inverse Laplace transform \cite{valko_comparison_2004,Mathematica} of $M^0$ and $M^1$ to obtain the first-passage time distribution:
\begin{align}
p (t) = p^0(t) + \nu p^1(t)
\label{eq:empirical_fptpdf}
\end{align}
neglecting, as usual, terms of higher order in $\nu$. In Fig.~\ref{fig:ouou_pdf}, we show the first-order corrected distribution compared to a numerical sampling of the probability function, $\tilde{p}(t)$, for values of $0 \leq \nu \leq 3.2$. The agreement for $\nu \lesssim 1$ is excellent, and plotting the rescaled deviation 
\begin{align}
\tilde{p}^1(t)|_{\nu} = \frac{ \tilde{p}(t)|_{\nu} - \tilde{p}(t)|_{\nu=0}} {\nu}
\label{eq:num_estimate_f1}
\end{align}
against the theoretical result of $p^1(y)$ (see inset of Fig.~\ref{fig:ouou_pdf}) shows systematic higher-order corrections consistent with our framework. Since for small $\nu$ the deviation to the undriven case is small, $\tilde{p} $ suffers from statistical noise, therefore we omit showing $\tilde{p}(\nu=0.1)$ in the inset.

These numerical results therefore confirm the analytically obtained first-order correction to the moment-generating function; consequently, the correction to all moments has been gained. As an illustration, we further show the first and second moment of the Ornstein-Uhlenbeck process driven by coloured noise in Fig.~\ref{fig:mom1_fpt_ouou} and Fig.~\ref{fig:mom2_fpt_ouou}. In analogy to the moment-generating function, we measure the mean and mean square first passage times $\tilde{\cT}_1 = \frac{1}{N}\sum_{i=1}^N \tilde\tau_i$, $\tilde{\cT}_2 = \frac{1}{N}\sum_{i=1}^N \tilde\tau_i^2$, which we assume to expand in $\nu$ as $\cT_1(\nu) = \cT^0_1 + \nu \cT_1^1 + \nu^2 \cT^2_1 + \ldots $ and $\cT_2(\nu) = \cT^0_2 + \nu \cT^1_2 + \nu^2 \cT_2^2 + \ldots $. The first-order corrections introduced are obtained by differentiation wrt $\bar{s}$
\begin{align}
    \cT^1_n =\left.\frac{\dd{} ^n}{\dd{(-\bar{s})^n}}\right|_{\bar{s}=0} M^1
    \label{eq:how_to_t_scaling}
\end{align}
using the result of \Eqref{eq:OUOU_M1}
which is performed by a computer algebra system and evaluated exactly. Due to their lengthiness, we do not give their full expression here. In order to numerically confirm these predictions, we measure the first order derivatives
\begin{align}
    \label{eq:def_cT1_ouou}
    \tilde{\cT}^1_1|_{\nu} &= \frac{\tilde{\cT}_1|_{\nu}  -\tilde{\cT}_1|_{\nu=0} }{\nu} \\
    \tilde{\cT}^1_2|_{\nu} &= \frac{\tilde{\cT}_2|_{\nu}  -\tilde{\cT}_2|_{\nu=0} }{\nu}
    \label{eq:def_cT2_ouou}
\end{align}
and compare it to the result obtained from \Eqref{eq:how_to_t_scaling}.
In Fig.~\ref{fig:mom1_fpt_ouou} and Fig.~\ref{fig:mom2_fpt_ouou}, we show the resulting moments of first-passage times obtained for fixed start position $x_0 = 0$ (at the minimum of the potential) but varied $x_1 \in [0.05,2]$. The figures show a clear agreement with the theoretical result and systematic deviations for larger $\nu$. Further, we observe that in this setting the coloured noise increases the mean-first passage time for smaller distances ($x_1 \lesssim 1.6$) and decreases it for larger distances. This also holds true for the mean squared first-passage time. This example therefore further illustrates that the effect of coloured driving (or memory) on the Langevin dynamics is highly non-trivial, yet our framework is able to capture this effect. The insets in both figures show the measured moments $\tilde{\cT}^1,\tilde{\cT}^2$.

\subsection{Active Thermal Brownian Motion on a ring (ATBM)}
\label{subsec:BMOU}
\begin{figure}[t]
\includegraphics[width=0.49\columnwidth]{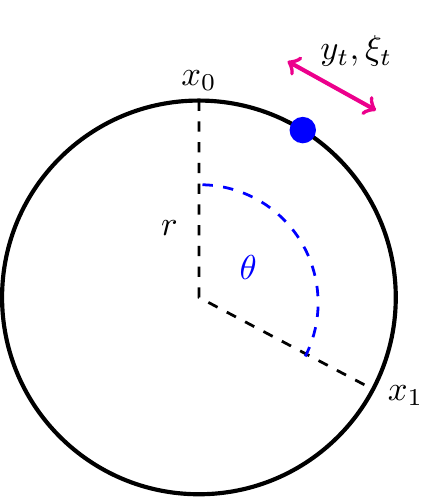}
\caption{A Particle on a circle of radius $r$ is driven by both white (thermal) and coloured (active) noise (\cf \Eqref{eq:bmou_sde}). We study the first passage time distribution from $x_0$ to $x_1$ as a function of the angle $\theta \in [0,2 \pi)$.}
\label{fig:atbm_ring}
\end{figure}

In this subsection, we consider the case of a Brownian particle at $x_t$ driven by coloured noise which is placed on a ring of radius $r$, thus satisfying periodic boundary conditions ($x \equiv x + 2 \pi r$).
The position of the particle satisfies the Langevin equation
\begin{align}
	\dot{x}_t = \xi_t + \varepsilon y_t
	\label{eq:bmou_sde}
\end{align}
with
\begin{align}
	\avg{\xi_{t_0} \xi_{t_1}} &= 2 D_x \delta(t_1-t_0) \\
	\yavg{y_{t_0} y_{t_1}} &= D_y \beta e^{-\beta|t_1-t_0|}.
\end{align}
We refer to this system as Active Thermal Brownian Motion  (ATBM) on a ring (see Fig.~\ref{fig:atbm_ring} for illustration).
In analogy to the previous subsection, we derive the correction to the moment generating function of the first-passage time distribution.

\subsubsection{From eigenfunctions to the moment generating function of first-passage times}

The Fokker Planck associated to the forward equation corresponding to Langevin Equation \eqref{eq:bmou_sde} is the Laplace equation with periodic boundary conditions $x \equiv x+ 2 \pi r$. Since its eigenfunctions are complex-valued square-integrable functions, we follow the identical steps of the framework introduced in Sec.~\ref{subsec:finding_expansion_terms} but replace the scalar-product in \Eqref{eq:orthogonality_condition} by its complex equivalent 
\begin{align}
\int \mathrm{d}{x}\, v_m^{*}(x) u_n(x) = \delta_{mn}.
\end{align} 
In what follows, we therefore consider the complex conjugate of the left eigenfunctions and obtain
	\begin{align}
		u_n(x) = \frac{1}{\sqrt{2\pi r}} e^{i \frac{n x}{r}} \\
		v_n^*(x) = \frac{1}{\sqrt{2\pi r}} e^{-i \frac{nx}{r}}
	\end{align}
	with corresponding eigenvalues
	\begin{align}
		\lambda_n = D_xr^{-2} n^2 \qquad (n \in \mathbb{Z}).
		\label{}
	\end{align}
	where by allowing $n \in \mathbb{Z}$, we enumerate the eigenfunctions efficiently (\cf \Eqref{eq:def_u_n}).
The eigenfunctions are conjugate to each other since the forward operator of simple diffusion, $\LL = D_x \partial_x^2$, is self-adjoint. From this follows that the noise-coupling matrix defined in \Eqref{eq:derivative_coupling_matrix},
\begin{align}
	\deriv_{mn} &= \int \mathrm{d} x \, v^*_{m} \partial_x u_n(x) \nonumber \\
	&=  \frac{1}{2\pi r}\int_{0}^{2\pi r} \dd{x}  \, i\frac{n}{r} e^{i(n-m)\frac{x}{r}} = i\frac{n}{r}\delta_{mn}
	\label{}
\end{align} 
is diagonal and purely imaginary. Because of scale-invariance and rotational symmetry, we simplify the following discussion by introducing the dimensionless angle
\begin{align}
    \theta \coloneqq \frac{x_1-x_0}{r},
\end{align}
where we restrict ourselves to $\theta \in [0,2\pi)$, and the diffusive timescale
\begin{align}
\label{eq:BM:def_alpha}
\alpha^{-1} = \frac{r^2}{D_x}
\end{align}
with which we rescale the Fourier-frequency $\omega$ to
\begin{align}
    \bar{s} = i\alpha^{-1} \omega,
\end{align}
again effectively evaluating the Laplace transform (\cf \Eqref{eq:Fourier_transform_functional}) of the respective probability densities.
With this simplified notation, the transition density to zeroth order reads
\begin{align}
	\transcoeff{0}{-i\bar{s}\alpha} &= \frac{1}{2\pi \alpha r}\sum_{n=-\infty}^{\infty} \frac{e^{-i n \theta }}{\bar{s} +  n^2 }  \nonumber \\ 
	& = \frac{1}{2\alpha r} \frac{\cosh\left( (\theta-\pi) \sqrt{\bar{s}} \right)}{ \sqrt{\bar{s}} \sinh\left( \pi \sqrt{\bar{s}}\right)},
	\label{eq:bmou_trans_zeroth}
\end{align}
and the return probability, setting $\theta = 0$, is
\begin{align}
	\returncoeff{0}{-i\bar{s}\alpha} &= \frac{1}{2\alpha r} \frac{\cosh\left( \pi \sqrt{\bar{s}} \right)}{ \sqrt{\bar{s}} \sinh\left( \pi \sqrt{\bar{s}}\right)}.
	\label{eq:bmou_return_zeroth}
\end{align}

Assuming the $y$-averaged moment generating function has an expansion of
\begin{align}
     &\avgfpt{s} 
     = M^{0, \rm{BM}}_{\frac{x_1 - x_0}{r}}\left( \frac{r^2 s }{D_x} \right)  + \underbrace{\frac{D_y \eps^2 \beta}{D_x \alpha}}_{\nu} M^{1, \rm{BM}}_{\frac{x_1 - x_0}{r}}\left(\frac{r^2 s}{D_x} \right) + \cO(\nu^2),
 \end{align} 
 where we use as in the Ornstein-Uhlenbeck case the dimensionless perturbative parameter
$     \nu= \frac{D_y \eps^2 \beta }{D_x \alpha}$ (here $\alpha$ has however a different definition than used in the OU case, see \Eqref{eq:BM:def_alpha}).
The zeroth order contribution is, using the classic result \Eqref{eq:FPTMGFclassic}, and the results in \eqref{eq:bmou_trans_zeroth}, \eqref{eq:bmou_return_zeroth},
\begin{align}
	M^{0, \rm{BM}}_{\theta} (\bar{s}) = \frac{\cosh\left( (\theta-\pi) \sqrt{\bar{s}}  \right)}{ \cosh\left(  \pi \sqrt{\bar{s}}  \right)}
	\label{eq:bm_classic_mgf}
\end{align}
which expands around $\bar{s}=0$ as
\begin{align}
	M^{0, \rm{BM}}_{\theta} (\bar{s}) =& 1+\left(\frac{\theta ^2}{2}-\pi  \theta \right) \bar{s} \\
	\nonumber&+\frac{1}{24} \left(\theta ^4-4 \pi  \theta ^3+8 \pi ^3 \theta \right) \bar{s} ^2 \\
	\nonumber&+\frac{1}{720} \left(\theta ^6-6 \pi  \theta ^5+40 \pi ^3 \theta ^3-96 \pi ^5 \theta \right) \bar{s} ^3+\ldots\nonumber
	\label{}
\end{align}
Being a moment generating function, the prefactors in front of $\bar{s}^n$ correspond to the (rescaled) moments of the first-passage time $\frac{(-1)^n}{n!} \avg{\tau_{x_0 \to x_1}^n} \left( \frac{D_x}{r^2} \right)^n$.

We turn to higher orders in $\eps$. To first order, the transition density 
\begin{align}
	\transcoeff{1}{-i\bar{s}\alpha,-i\bar{\beta}\alpha} &= \frac{1}{2\pi  \alpha^2 r^2}\sum_{n=-\infty}^{\infty} \frac{i ne^{-i n \theta}}{(\bar{s}+\bar{\beta}+n^2)(\bar{s}+n^2)} \nonumber\\
	&=  \frac{1}{2\pi  \alpha^2 r^2}\sum_{n=0 }^{\infty} \frac{ -2n{\sin{ n \theta}}}{(\bar{s}+\bar{\beta}+n^2)(\bar{s}+n^2)}
	\label{}
\end{align}
Here, we introduced  as before $\bar{\beta} = \alpha^{-1}\beta$
as the dimensionless correlation time-scale of the coloured noise. The function $\hat{R}^1(-i\bar{s}\alpha,-i\bar{\beta}\alpha)$, which corresponds to $\hat{T}^1(-i\bar{s}\alpha,-i\bar{\beta}\alpha)$ evaluated at $\theta=0$, vanishes. This implies that the contribution of integrals $({\rm II})$ and $({\rm III})$, as given in Eq.~\eqref{eq:diagram_II} and \eqref{eq:diagram_III}, vanish, leaving only $({\rm I})$ and $({\rm IV})$ as correction terms.

To second order, the transition density is
\begin{widetext}
\begin{align}
	\transcoeff{2}{-i\bar{s}\alpha,-i\bar{\beta}\alpha,i\bar{\beta}\alpha} &= -\frac{1}{\pi \alpha^3 r^3}\sum_{n=-\infty}^{\infty} \frac{n^2e^{-i n \theta}}{(\bar{s}+n^2)^2(\bar{s}+\bar{\beta}+n^2)} \nonumber \\
	&= - \frac{1}{\pi \alpha^3 r^3} \sum_{n=-\infty}^{\infty} \left[ \frac{1}{{\bar\beta}^{2}} \frac{n^2 e^{-in\theta}}{\bar{s} + \bar{\beta} + n^2} - \frac{1}{ {\bar{\beta}}^{2}} \frac{n^2e^{-i n \theta}}{\bar{s} + n^2} + \frac{1}{ {\bar{\beta}} } \frac{n^2e^{-in \theta}}{(\bar{s}+ n^2)^2} \right] \nonumber \\
	& = \frac{2}{\alpha^2 r^2} \cdot \frac{1}{ {\bar{\beta}}^2 } \partial_{\theta}^2 \left[ \transcoeff{0}{\bar{s}+ \bar{\beta}}  - \transcoeff{0}{\bar{s}} - \bar{\beta} \partial_{\bar{s}} \transcoeff{0}{\bar{s}}\right] \nonumber \\
	&= \frac{2}{\alpha^3 r^3}\cdot \frac{1}{4{\bar{\beta}}^2 \sqrt{\bar{s}}} \left[\frac{\cosh\left( (\theta-\pi)\sqrt{\bar{s}} \right)}{\sinh\left( \pi \sqrt{\bar{s}} \right)} \left( \pi \bar{\beta} \sqrt{\bar{s}} \coth\left( \pi \sqrt{\bar{s}} \right)-2\bar{s} -\bar{\beta} \right) \right. \nonumber \\
	&\qquad \qquad \qquad +\sqrt{\bar{s}} \left. \left\{ 2 \sqrt{\bar{\beta}+\bar{s}} \frac{\cosh\left( (\theta-\pi)\sqrt{\bar{s}} \right)}{\sinh(\pi\sqrt{\bar{\beta} + \bar{s})}} + \bar{\beta}(\pi-\theta)\frac{\sinh\left( (\theta-\pi)\sqrt{\bar{s}} \right)}{ \sinh(\pi \sqrt{\bar{s}}}   \right\} \right]
\end{align}
where $\partial_{\theta} = r \partial_{x_1}$. Setting $\theta = 0$, one obtains the second order response of the return probability,
	\begin{align}
	&\returncoeff{2}{-i\bar{s}\alpha,-i\bar{\beta}\alpha,i\bar{\beta}\alpha} = \frac{1}{2 \alpha^3 {\bar{\beta}}^2 r^3} \left[ \pi  \bar{\beta}  \coth ^2\left(\pi  \sqrt{\bar{s} }\right)-\frac{(\bar{\beta} +2 \bar{s} ) \coth \left(\pi  \sqrt{\bar{s} }\right)}{\sqrt{\bar{s} }} + 2 \sqrt{\bar{\beta} +\bar{s} } \coth \left(\pi  \sqrt{\bar{\beta} +\bar{s} }\right)-\pi  \bar{\beta} \right],
	\end{align}
Inserting these quantities into the FPT correction \eqref{eq:fpt_correction_general}, gives 
\begin{align}
	&M(s=\alpha \bar{s}) =\frac{\cosh\left( (\theta - \pi)\sqrt{\bar{s}} \right)}{ \cosh\left( \pi\sqrt{\bar{s}}  \right)} 
	+ \frac{D_y \varepsilon^2}{2 D_x} \cdot
	\frac{\sqrt{\bar{s}}\tanh\left( \pi \sqrt{\bar{s}} \right)}{{\bar{\beta}}^2} \left[ \frac{\cosh\left( (\theta - \pi)\sqrt{\bar{\beta}+\bar{s}} \right)}{\sinh\left( \pi\sqrt{\bar{\beta}+\bar{s}} \right)}2\sqrt{\bar{\beta}+\bar{s}}\right.
		\nonumber \\
	& \qquad + \left.\frac{\cosh\left( (\theta-\pi) \sqrt{\bar{\beta}+\bar{s}} \right)}{ \cosh\left( \pi \sqrt{\bar{s}} \right)} \left( \pi\bar{\beta}-2\sqrt{\bar{\beta}+\bar{s}} \coth\left( \pi \sqrt{\bar{\beta}+\bar{s}} \right) \right)+ \frac{\sinh\left( (\theta-\pi)\sqrt{\bar{s}} \right)}{\sinh\left( \pi\sqrt{\bar{s}} \right)}\bar{\beta}\left( \pi-\theta \right) \right] + \cO(\eps^4)
	\label{eq:bmou_full_mgf}
\end{align}
This expression is the \emph{full moment generating function up to second order in $\varepsilon$} (first order in $\nu$) in elementary functions. In line with our previous notation, we identify the dimensionless scaling functions
\begin{align}
  M^{0, \rm{BM}}_{\theta} (\bar{s})  &= \frac{\cosh\left( (\theta - \pi)\sqrt{\bar{s}} \right)}{ \cosh\left( \pi\sqrt{\bar{s}}  \right)} \label{eq:mgf0_bmou} \\
 M^{1, \rm{BM}}_{\theta} (\bar{s}) &= \frac{\sqrt{\bar{s}}\tanh\left( \pi \sqrt{\bar{s}} \right)}{2 {\bar{\beta}}^2}\left[ \frac{\cosh\left( (\theta - \pi)\sqrt{\bar{\beta}+\bar{s}} \right)}{\sinh\left( \pi\sqrt{\bar{\beta}+\bar{s}} \right)}2\sqrt{\bar{\beta}+\bar{s}}\right.   \label{eq:mgf1_bmou} \nonumber	\\
   &\left.+ \bar{\beta}  (\pi -\theta ) \frac{\sinh \left((\theta -\pi ) \sqrt{\bar{s} }\right)}{\sinh\left(\pi  \sqrt{\bar{s} }\right)}+\left(\pi  \bar{\beta} -2 \sqrt{\bar{\beta} +\bar{s} } \coth \left(\pi  \sqrt{\bar{\beta} +\bar{s} }\right)\right) \frac{\cosh \left((\theta -\pi ) \sqrt{\bar{s} }\right)}{\cosh \left(\pi  \sqrt{\bar{s} }\right)} \right]
\end{align}
which together form the first-order correction of the moment-generating function
\begin{align}
 M(s) =M^{0, \rm{BM}}_{\frac{x_1-x_0}{r}} \left( \frac{r^2 s}{D_x} \right) + \frac{D_y  \eps^2 \beta}{D_x \alpha}M^{1, \rm{BM}}_{\frac{x_1-x_0}{r}} \left( \frac{r^2 s}{D_x} \right) +\cO(\eps^4)\ .
\end{align}
An expansion around $\bar{s} = 0$ gives corrections to all moments. The correction to first order in $\nu$ of the  mean first-passage time over an angle of $\theta$, $\avg{\tau_{0\to \theta}}$, for instance reads
\begin{align}
	\alpha\avg{\tau_{0, \theta}} = \underbrace{\pi \theta - \frac{\theta^2}{2}}_{\cT^1_0(\theta)}- \nu \underbrace{\frac{1}{2 {\bar{\beta}}^{3/2}} \left( \sqrt{\bar{\beta}}(2\pi-\theta)\theta- 2\pi \coth\left( \pi \sqrt{\bar{\beta}} \right)+2\pi\frac{\cosh\left( \sqrt{\bar{\beta}}(\theta - \pi) \right)}{\sinh(\pi \sqrt{\bar{\beta}})} \right)  }_{=: \cT^1_1(\theta, \bar{\beta})} + \cO(\nu^2)
	\label{eq:mfpt_full}
\end{align}
\end{widetext}
where we indicated that the result can be written as the classical contribution ($\cT^1_0$), plus the dimensionless perturbative coefficient $\nu$ times a dimensionless scaling function $\cT^1_1(\theta, \bar{\beta})$. By successive derivation, any higher order moment may be obtained from \Eqref{eq:bmou_full_mgf}.

In the limit of an infinitely large radius $r \to \infty$ one obtains, under suitable rescaling, the moment generating function for an active thermal Brownian motion on the real line. 
We have confirmed that the perturbative approximation \Eqref{eq:bmou_full_mgf} agrees, up to second order in $\eps$, with the analytic solution for a run-and-tumble process subject to additional white noise in the limit of small tumble noise which has been calculated in \cite{malakar_steady_2018}. 

\subsubsection{Numerical Validation}
\label{subsubsec:numerical_validation_bmou}
\begin{figure}
\includegraphics[width=1.0\columnwidth]{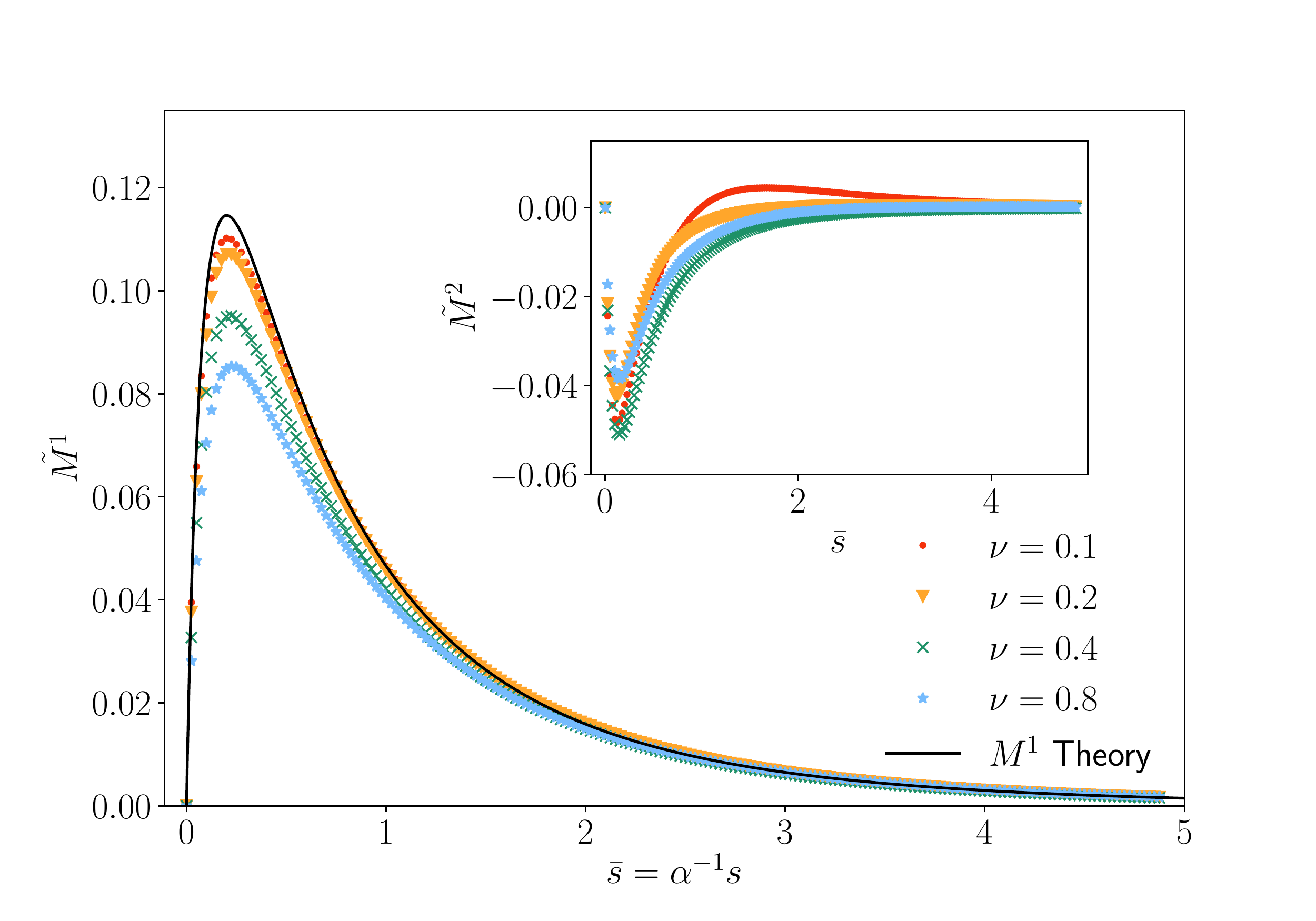}
\caption{\textbf{Numerical validation of first order correction to FPT moment generating function} $M^1$ of ATBM (\cf \Eqref{eq:bmou_sde} and \secref{subsec:BMOU} for discussion) for various values of $0.1 \leq \nu \leq 0.8$. The moment generating functions were sampled for $x_0=0, x_1 = \pi$, $D_x = 1$, $\alpha = 1$, $D_y = 1$, $\beta = \frac{1}{2}$ and $\eps$ suitably chosen to fix $\nu = D_yr^2  \beta  \eps^2/(D_x^2)$ (\textbf{plot marks}). For small values of $\nu$ agreement with theoretical first-order correction (\textbf{black line}) is very good. For larger values of $\nu$ the deviation  increases. The rescaled deviations, $\tilde{M}_2$ (see \Eqref{eq:def_cM2_bmou}), (\textbf{inset}) collapse and thus confirm that these deviations are systematic higher-order corrections. See \eqref{subsubsec:numerical_validation_bmou} for further results and discussion.}
\label{fig:mgf_bmou}
\end{figure}

In order to validate the analytic result of the moment generating function of first-passage times \Eqref{eq:bmou_full_mgf} and the mean first-passage time \Eqref{eq:mfpt_full}, we follow the same steps as in the previous section \ref{subsubsec:numerical_validation_ouou}. Using Monte-Carlo simulations, we sample the first-passage times $\tilde{\tau}_i$ of the integrated stochastic equation \eqref{eq:bmou_sde}. To validate the moment-generating function, we average over $N\simeq 10^6$ to $10^7$ iterations to sample
\begin{align}
    \tilde{M} = \frac{1}{N}\sum_{i=1}^N e^{-s\tilde{ \tau}_i}
\end{align}
for various values of $\nu$ and across $\target-\start \in (0,\pi]$ (with $r=1$). Again, symbols with a tilde denote quantities which are measured numerically. To validate the theoretically predicted first-order correction \Eqref{eq:mgf1_bmou}, we introduce the numerically measured corrections:
\begin{align}
    \tilde{M}^1|_{\nu}&= \frac{\tilde{M}|_{\nu}-\tilde{M}_{\nu=0}}{\nu} 
    \label{eq:def_cM1_bmou}\\
    \tilde{M}^2|_{\nu} &= \frac{\tilde{M}^1|_{\nu}-M^1}{\nu} ~.
    \label{eq:def_cM2_bmou}
\end{align}
In Fig.~\ref{fig:mgf_bmou}, we show the analytic result $M^1$ (\cf \Eqref{eq:mgf1_bmou}) and numerically obtained $\tilde{M}^1|_{\nu}$ (\cf \Eqref{eq:def_cM1_bmou}) for various values of $0 \leq \nu \leq 0.8$. The agreement is again excellent, and the discrepancy between simulated result and theoretical first order correction grows, as expected, with larger values of $\nu$. The rescaled discrepancy, $\tilde{M}^2$, to leading order the second-order correction $M^2$ (\cf \Eqref{eq:def_cM2_bmou}), is plotted in the inset and collapses, indicating that the discrepancy is systematic and confirming the validity of the result in \Eqref{eq:mgf1_bmou}.  By numerically computing the inverse Laplace transform, we obtain the perturbative correction to the first-passage time distribution (\cf \Eqref{eq:empirical_fptpdf}) which in Fig.~\ref{fig:bmou_pdf} is compared to the numerically sampled distribution.
The inset shows the numerical estimate of the first order correction (\cf \Eqref{eq:num_estimate_f1}) which agrees well for small $\nu$. For larger values of $\nu$, second-order corrections play an increasing role as indicated in Fig.~\ref{fig:mgf_bmou}.

\label{subsec:atbmnumerics}
\begin{figure*}
    \subfloat[\textbf{Correction to mean first passage time of ATBM on a ring} (\cf \Eqref{eq:bmou_sde}) as obtained from \Eqref{eq:def_cT1_ouou} versus target position $x_1$ and various values of $\nu$ (\textbf{plot marks}). This is  compared to theoretical result of \Eqref{eq:mfpt_full} (\textbf{solid black line}). The \bf{inset} shows the measured mean first passage time versus a varying target position $x_1$ and different values of $\nu$. ]{\includegraphics[width=\columnwidth]{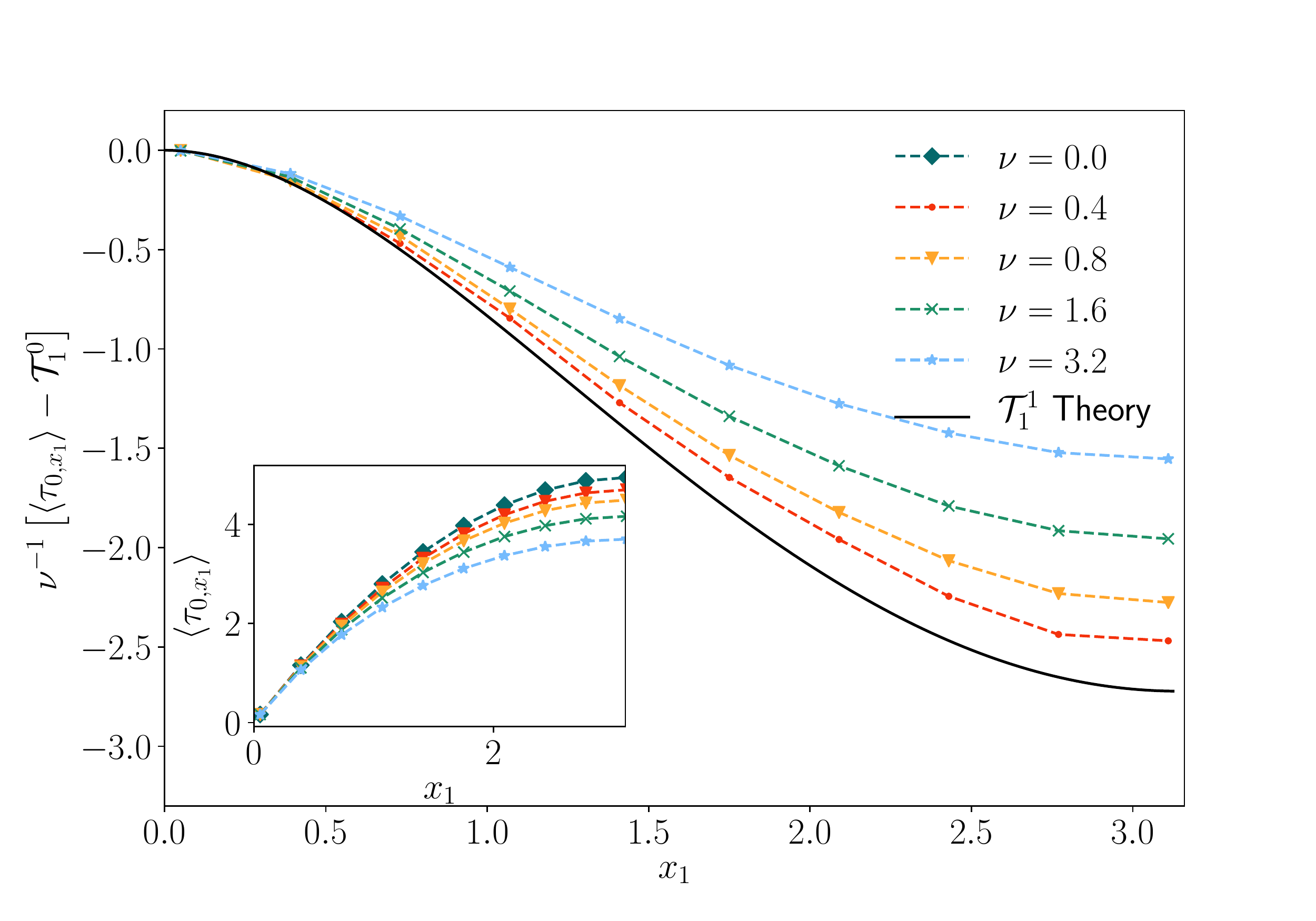}\label{fig:t1_bmou}}
    \quad
    \subfloat[\textbf{Correction to mean squared first passage time of ATBM on a ring} (\cf \Eqref{eq:bmou_sde}) as obtained from \Eqref{eq:def_cT2_ouou} versus target position $x_1$ and various values of $\nu$ (\textbf{plot marks}) . This is  compared to theoretical result of twice differentiating \Eqref{eq:mgf1_bmou} (\textbf{solid black line}). The \bf{inset} shows the measured mean squared first passage time versus a varying target position $x_1$ and different values of $\nu$. ]{\includegraphics[width=\columnwidth]{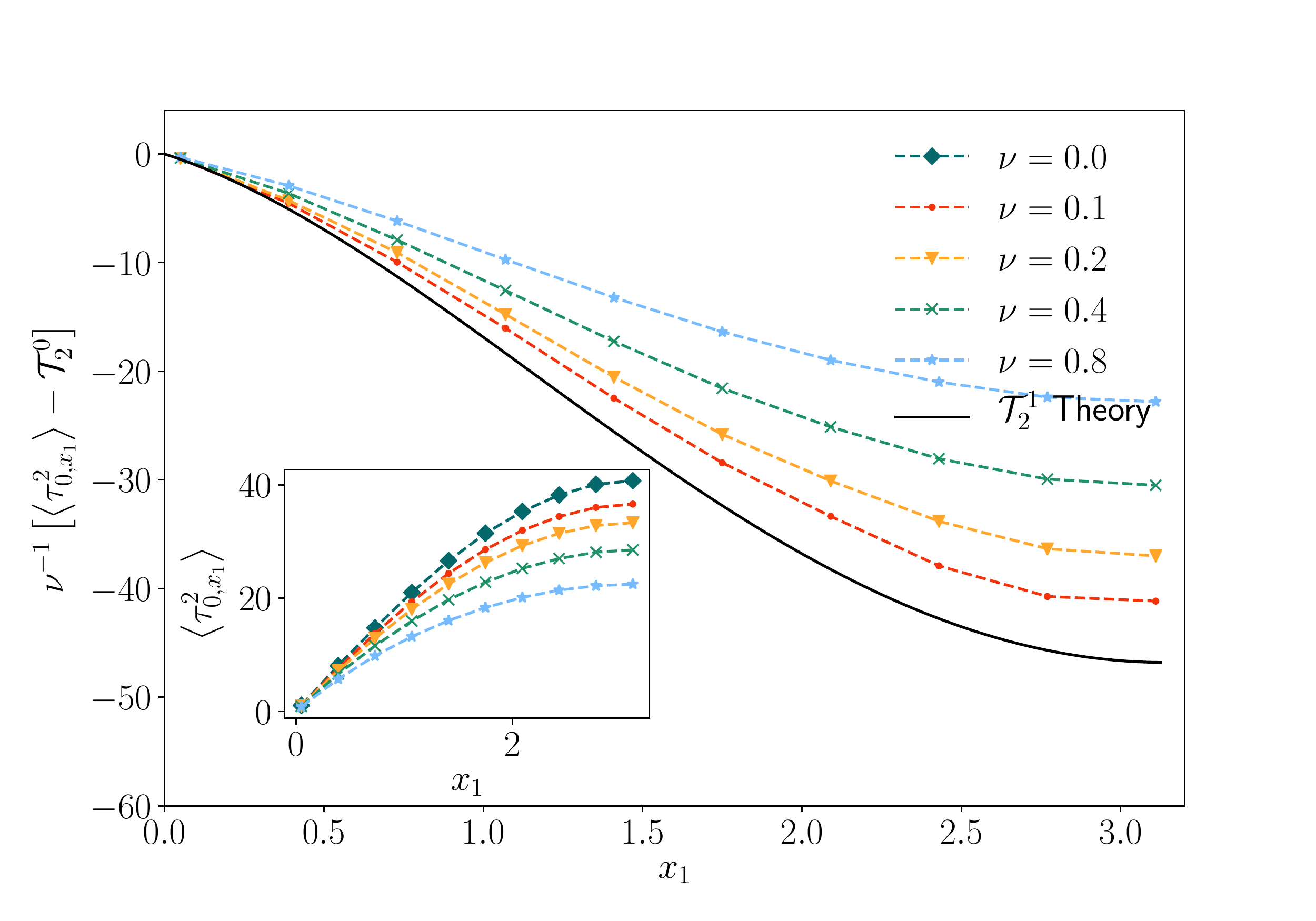}    \label{fig:t2_bmou}}
    \label{}
    \caption{First order correction to first and second moment of first-passage times of ATBM with periodic boundary conditions. Simulation parameters are $x_0 = 0, D_x = 1, r=1, D_y = 1, \beta = \frac{1}{2}$ and $\eps$ suitably chosen to fix $\nu = D_y r^2 \beta \eps^2/(D_x^2)$.}
\end{figure*}

In Figs.~\ref{fig:t1_bmou} and \ref{fig:t2_bmou}, we show the first and second moment of first-passage times and how their deviations to the $\nu=0$ case is captured by the first-order correction obtained using \Eqref{eq:def_cT1_ouou} and \Eqref{eq:def_cT2_ouou} with the result of \Eqref{eq:mgf1_bmou}. For the first and second moment, the agreement is again excellent, showing that the correction induced by the active driving noise is accurately captured to leading order. The insets of Figs.~\ref{fig:t1_bmou} and \ref{fig:t2_bmou} show the respective moments of the FPT for various $\nu$, indicating the systematic decrease for increased values of $\nu$.

\subsection{Limit Cases}
\label{subsec:limit_cases}
The framework we introduce here allows to study coloured driving noises at any correlation time $\beta^{-1}$. In particular, this includes two limit cases of $\beta \to 0$ and $\beta \to \infty$. For appropriate re-scaling of $D_y$, the former limit corresponds to a particular quenched disorder model, and the latter to additional white noise. In what follows we discuss these limit cases in more detail.

\subsubsection{The white noise limit}
For very small autocorrelation times $\beta^{-1}$ the driving noise $y_t$ appears more and more as white noise. The correlator of $y_t$ tends towards
\begin{align}
    \yavg{y_{t_0}y_{t_1}} = D_y \beta e^{-\beta|t_1-t_0|} \to  2 D_y \delta(t_1-t_0), \qquad \beta \to \infty.
\end{align}
In this limit, the driving noise features in the Langevin equation \eqref{eq:driven_langevin} as additional white noise and is absorbed as
\begin{align}
    \dot{x} = -V'(x_t) + \sqrt{2\left(D_x + \eps^2 D_y  \right) }\xi,
\end{align}
such that effectively the diffusion constant is shifted by $D_x \mapsto D_x + \eps^2 D_y$. In the white noise limit, the theory is Markovian and Eq.~ \eqref{eq:FPTMGFclassic} may be applied using the shifted diffusivity to obtain exact results.The perturbation theory presented here then correspond to an expansion of the Markovian result in a perturbation to the diffusion constant $D_x$; and as a result the following identity must hold:
\begin{align}
   \lim_{\beta \to \infty} \beta M^1_{x_0, x_1}(s) =\alpha D_x \frac{\partial }{\partial D_x} M^0_{x_0, x_1}(s) .
\end{align}

One can verify that this relation is indeed satisfied for the Brownian motion case (Eqs. \eqref{eq:mgf0_bmou} and \eqref{eq:mgf1_bmou}) as well as the Ornstein-Uhlenbeck case (Eqs. \eqref{eq:OUOU_M0} and \eqref{eq:OUOU_M1}).

\subsubsection{Quenched Disorder limit}
In the opposite limit of $\beta \to 0$, provided $D_y\beta=w^2$ remains fixed, the driving noise ``freezes'' to a random constant since
\begin{align}
    \yavg{y_{t_0}y_{t_1}} = D_y \beta e^{-\beta|t_1-t_0|} \to w^2, \qquad \beta \to 0.
\end{align}
Effectively, the Langevin equation \eqref{eq:driven_langevin} therefore turns into 
\begin{align}
    \dot{x} = -V'(x_t) +\xi_t +  v
    \label{eq:quenched_langevin}
\end{align}
 where $v$ is a constant driving velocity which is normal distributed according to $v \sim \mathcal{N}(0, \eps^2 w^2)$. The driving noise average $\yavg{e^{-s \fptxy}}$ then corresponds to a quenched average over the ensemble of normal distributed velocities $v$. If we treat  $M^1$ as a functional of the potential $V(x)$ in which the particle is embedded, then  formally
 \begin{align}
     &\frac{w^2 \eps^2}{D_x \alpha}\lim_{\beta \to 0} M^1_{x_0, x_1}(s;[V(x)]) \\
     &= \frac12 \underbrace{\eps^2 w^2}_{=\yavg{v^2}} \left. \partial_v^2 \right|_{v=0} M_{x_0, x_1}(s;[V(x) + vx]).
     \label{eq:quenched_mgf_are_the_same}
 \end{align}
Our framework therefore predicts the first-order correction in $\yavg{v^2}$.

 \paragraph*{Example: Brownian Motion with periodic boundary conditions and a random drift}
 For Brownian Motion with periodic boundary conditions, as studied in Sec.~\ref{subsec:BMOU}, one can compute the moment generating function of first-passage times for a particular fixed drift $v$ exactly (see Appendix \ref{app:bm_with_drift} and in particular Eq.~\eqref{eq:quenched_bmou_MGF_exact} for the result. We could not find this result elsewhere in the literature.) On expanding this result in orders of the drift $v$ and averaging $v^2$ over its distribution $\mathcal{N}(0,\eps^2 w^2)$ we obtain a resulting quenched average approximation in orders of $\yavg{v^2}$, ${\yavg{\avgfpt{s;v}}} = {{\avgfpt{s;v=0}}} + \frac12 {{\left. \partial_v^2 \right|_{v=0}\avgfpt{s;v}}}\cdot \yavg{v^2} + ...$.
 When employing our framework and letting $\beta \to 0$ in our general result \Eqref{eq:bmou_full_mgf} we recover precisely $\frac12 {{\partial_v^2 |_{v=0} \fpt{s;v}}}$. The necessary calculations are given in App.~\ref{app:equivalent_quenched_averages} and show that this is indeed the case. By way of this relation, our framework for instance returns the correction to the mean-first passage time of a Brownian motion with quenched disordered drift to first order in $\nu$ as (compare to \Eqref{eq:mfpt_full})
 \begin{align}
     \alpha \avg{\yavg{\tau_{0,\theta}}} = \frac{\theta \left( 2\pi -\theta \right)}{2} - \frac{\eps^2 w^2}{\alpha^2 r^2 }\cdot \frac{\theta ^2 (\theta -2 \pi )^2}{24 } + ...
     \end{align}
     such that quenched disorder lowers the mean first-passage time for any choice of parameters. Further, for the mean squared first-passage time we obtain
     \begin{align}
    & \alpha^2 \avg{\yavg{\tau_{0,\theta}^{2}}} =\frac{1}{12} \theta \left(\theta-2\pi\right)\left( \theta - 2\varphi \pi\right) \left( \theta - 2\pi + 2\varphi \pi\right) \nonumber \\
         &+\frac{\eps^2 w^2}{\alpha^2 r^2 }\frac{\theta(\theta-2\pi)(\theta-\pi(1+\psi^+))(\theta-\pi(1-\psi^+))}{120} \nonumber \\
         &\qquad \cdot (\theta-\pi(1+\psi^-))(\theta-\pi(1-\psi^-)),
     \end{align}
     where $\varphi = \frac{1+\sqrt{5}}{2}$ is the golden ratio, and $\psi^{\pm}= \frac{1\pm\sqrt{10}}{\sqrt{3}}$.

\section{\label{sec:discussion} Conclusion}

In the present work, we introduce a perturbative approach to study the first-passage time distribution of stochastic processes which are driven both by white and coloured noise. This class of stochastic processes lies at the heart of the study of active particles in a thermal environment. The activeness, which can either represent self-propulsion or hidden degrees of freedom, is modelled by coloured noise $y_t$ with characteristic timescale $\beta^{-1}$ while the thermal bath is modelled by white noise $\xi_t$ with diffusion constant $D_x$. The expansion parameter in which the perturbation takes place is a dimensionless quantity, $\nu$, which indicates how strong the fluctuations of the active noise are in comparison to the strength of thermal fluctuations. We study the regime in which $\nu$ is small.

Setting out from a renewal equation which gives the moment generating function of first-passage times, we employ a functional expansion to obtain its perturbative corrections. This key equation \eqref{eq:main_eq_to_solve}  stands at the centre of this work.  In order to solve it perturbatively, one needs to calculate the expansion terms (\cf Eqs.~\eqref{eq:functional_expansion_second_exposition} and \eqref{eq:expansion_Rminus1}) which involve the eigenfunctions of the Fokker Planck operator associated to the non driven process (\cf \Eqref{eq:def_u_n}). To first order in $\nu$, we obtain an analytic result of the moment-generating function in terms of the associated eigenfunctions (cf.~\Eqref{eq:fpt_correction_general}). Higher order contributions can be obtained  by further iterating the steps outlined in Sec.~\ref{subsec:finding_expansion_terms}.

To illustrate  the capabilities of our framework, we study two systems.
First, we consider an active thermal particle  in a harmonic potential, the Active Thermal Ornstein-Uhlenbeck Process. In Sec.~\ref{subsec:OUOU_eigenfunctions}, we calculate all necessary response functions to find the first order correction to the moment-generating function of first-passage times (cf.~\Eqref{eq:OUOU_M1}). By taking derivatives, we could in principle obtain closed form expressions for the first order correction to any moment of the first-passage time distribution. We confirm these analytical results by numerical simulations. Sampling the experimental moment generating function, we obtain an excellent agreement with the first-order correction (see Fig.~\ref{fig:ouou_m1}). For larger values of $\nu$, the perturbative parameter, the deviations systematically indicate higher-order corrections. Further, we compare the theoretically predicted correction to the first two moments of the first-passage times to numerical results (see Figs.~\ref{fig:mom1_fpt_ouou} and \ref{fig:mom2_fpt_ouou}) which are in excellent agreement.
Secondly, we study Active Thermal Brownian Motion on a ring (see Sec.~\ref{subsec:BMOU}). Again, we illustrate our framework by finding the first-order correction to the moment generating function (\cf \Eqref{eq:bmou_full_mgf}). Numerical simulations show excellent agreement and systematic higher-order corrections (see Fig.~\ref{fig:mgf_bmou}). Both first and second moment of the first-passage time are obtained from \Eqref{eq:bmou_full_mgf} and show good agreement with numerical simulations.

Further, since the perturbation theory we present makes no assumption on $\beta^{-1}$, we are able to recover the limiting cases for $\beta \to 0$ and $\infty$, respectively. The case of $\beta \to 0$ is of particular interest since it recovers  quenched disorder averages over processes with additional fixed and normal distributed drift (see Sec.~\ref{subsec:limit_cases}).

The framework requires to find the eigenfunctions of a differential operator, and to express all transition and return densities as sums over these eigenfunctions. This often requires certain calculations that for more unusual eigenfunctions may be difficult to perform.

Our approach further allows for the presence of an external potential provided the associated differential operator (\Eqref{eq:forward_equation}) can be diagonalised. This significantly extends the range of systems our framework can be applied to.  In this work, we focused on Fourier-modes and Hermite-polynomials which are suitable for flat and harmonic potentials. It is, however, also possible to study piece-wise combinations of the potentials using these eigenfunctions. This may be relevant when studying bi-stable processes for instance.
Further, as long as \Eqref{eq:forward_equation} can be diagonalised,
 this framework also allows for a space-dependent thermal diffusivity by letting $D_x = D(x)$. 
For future work, for instance, it would be interesting to study first-passage time behaviour of particles at the boundary between two heat baths at different temperature (\eg $D(x) = D_0 + \operatorname{sgn}(x)\Delta D$).

Moreover, the functional expansion in $\hat{y}(\omega)$ (\cf \Eqref{eq:rho_functional_expansion}) drastically simplifies in the case of $y_t$ being a periodic driving force. This framework therefore would not only be able to capture stochastic $y$, but also oscillating deterministic driving forces. This will possibly be addressed in future work.

To first order in $\nu$, the corrections, as given in Eqs.~\eqref{eq:diagram_I} to \eqref{eq:diagram_III}, involve simple complex integrals which can be solved by the residue theorem. For higher-order corrections, however, the integration runs over more than one free internal variable and will require more work. This corresponds to the problems of typical Feynman-diagrams of higher order in statistical field theories which often involve non-trivial integrals. To study higher orders, field theory therefore would provide the necessary toolbox to solve the required correction terms. 
Already, the general result obtained in \Eqref{eq:fpt_correction_general}  can be analogously derived  using field-theoretic methods (see \cite{Walter2020}).

In conclusion, by casting a broad class of non-Markovian processes into a perturbative language, our framework will prove itself useful in tackling a diverse range of future challenges.

\acknowledgements
BW and GP would like to thank the Francis Crick Institute for hospitality and Andy Thomas for invaluable computing support. BW thanks Kay Wiese, Grigorios Pavliotis, Gregory Schehr and Raphael Voituriez for insightful discussions. GS was supported by the Francis Crick Institute which receives its core funding from Cancer Research UK (FC001317), the UK Medical Research Council (FC001317), and the Wellcome Trust (FC001317).

\renewcommand{\emph}[1]{\textit{#1}}
\bibliography{main}

\appendix
\renewcommand{\emph}[1]{%
  \uline{\phantom{#1}}%
    \llap{\contour{white}{#1}}%
    }
 \begin{widetext}

\section{Notations}
\label{appendix:notations}
For reference, we list the most commonly used notations in Table ~\ref{tab:symbols}.
\begin{table}[h!]
\begin{tabular}{l l}
\textsc{Symbol} & \textsc{Description}\\
\hline
$p_{x_0, x_1}(t_0, t_1)$ & First-passage time probability density at $(x_1, t_1)$ starting from $(x_0, t_0)$ \\
$\hat{p}_{x_0, x_1}( \omega)$ & Fourier transform (characteristic function) of FPT density \\
$M_{x_0, x_1}( s)$ & Laplace transform (moment generating function) of FPT density \\
$T_{x_0,x_1}(t_0, t_1) $ & Probability density of progress from $(x_0,t_0)$ to $(x_1, t_1)$ \\
$\hat{T}_{x_0, x_1}(\omega) $ & Fourier transform of transition probability (\cf \Eqref{eq:diagonal_function}) \\
$R_{x_1}(t_0, t_1)$ & Probability density of return at $x_1$ at time $t_1$ starting at $t_0$ \\
$\hat{R}_{x_1}(\omega) $ & Fourier transform of return probability (\cf \Eqref{eq:diagonal_function}) \\
$ \hat{T}^{(n)}_{x_0,x_1}(\omega_1,\ts_1,...,\ts_n)$ & Expansion functionals of transition probability in $\eps y_t$ (\cf \Eqref{eq:def_t_expansion})\\  
$ \hat{R}^{(n)}_{x_1}(\omega_1,\ts_1,...,\ts_n)$ & Expansion functionals of return probability in $\eps y_t$ (\cf \Eqref{eq:def_r_expansion})\\  
$M_{x_0,x_1}^0, M_{x_0,x_1}^1 ,...$& $y_t$-Averaged coefficients to moment generating function $M_{x_0, x_1}( s)$ in $\nu$-expansion \\
\end{tabular}
\caption{Overview of symbols used for various probability densities and their functional transforms. When possible, the lower subscripts $x_0$, $x_1$ are ommitted.}
\label{tab:symbols}
	\end{table}
\section{First passage times of Brownian Motion on a ring with drift}
\label{app:bm_with_drift}
We calculate the first-passage time distribution of a Brownian Motion with drift $v$ on a ring of radius $r$  departing from the methods outlined in \cite[Chp. V]{Cox1965} although the explicit formula is not given there. Instead of considering the first passage event of transition $x_0 \to x_1$, we calculate the exit probability of a Brownian Motion on the real line over the absorbing boundaries at $x_1$ and $x_1 - 2\pi r$ where without loss of generality we chose $x_1 - 2\pi r < x_0 < x_1$. The transition density $\avgtrans{t} \equiv T_{x_0, x_1}( t)$ satisfies the Fokker-Planck Equation
\begin{align}
	\partial_t \avgtrans{t} = D_x\partial_{x_1}^2\avgtrans{t} - v \partial_{x_1} \avgtrans{t}
	\label{app:eq:fpe_bm_on_ring}
\end{align}
and the Kolmogorov backward equation
\begin{align}
	\partial_t \avgtrans{t} = D_x\partial_{x_0}^2\avgtrans{t} + v \partial_{x_0} \avgtrans{t}
	\label{app:eq:kbe_bm_on_ring}
\end{align}
with absorbing boundary conditions for $t\geq0$
\begin{align}
	T_{x_0,x_1}( t) &= 0\\
	T_{x_0, x_1-2\pi r}(t) &= 0
\end{align}
and initial condition
\begin{align}
	T_{x_0, x}(  t=0) = \delta(x - x_0)
	\label{}
\end{align}
The net flux of the surviving probability over the boundary is related to the first-passage time distribution via
\begin{align}
 p_{x_0, x_1}( t) = -\partial_t \int_{x_1 - 2\pi r}^{x} \dd{x} T_{x_0,x}(t).
\end{align}
Using the backward equation \eqref{app:eq:kbe_bm_on_ring}, and applying a further derivative in time one finds
\begin{align}
\partial_t  p_{x_0, x_1}( t) &= - (D_x \partial_{x_0}^2 + v \partial_{x_0})  \partial_t  \int_{x_1 - 2\pi r}^{x_1}\dd{x} T_{x_0,x}(t) \\
&=(D_x \partial_{x_0}^2 + v \partial_{x_0})p_{x_0, x_1}( t) . \label{appeq:fpt_laplace1}
\end{align}
At the boundary, the first-passage time distribution is given by
\begin{align}
p_{x_1, x_1}( t)= p_{x_1-2\pi r, x_1}(t) = \delta(0^+)
\end{align}
since the particle is immediately absorbed, while at time $t=0$ for $x_0$ away from the boundary
\begin{align}
p_{x_0, x_1}( t=0) = 0 
\end{align}
since the particle has not yet begun to diffuse.
Under Laplace transform in $t$, the differential equation \eqref{appeq:fpt_laplace1}, together with these boundary and initial conditions returns
\begin{align}
\begin{cases}
s p_{x_0, x_1}( s) = (D_x \partial_{x_0}^2 + v \partial_{x_0})p_{x_0, x_1}( s) & x_1 - 2\pi r < x_0 < x_1 \\
p_{x_0, x_1}( s) =p_{x_0, x_1-2\pi r}( s)  = 1 & x_0 = x_1-2\pi r \text { or } x_0 = x_1.
\label{appeq:fpt_laplace}
\end{cases}
\end{align}
The ordinary differential equation (away from the boundary) is solved by the exponential ansatz
\begin{align}
	M_{x_0, x_1}(  s) = A e^{\omega_1 x_0} + B e^{\omega_2 x_0}
\end{align}
Inserting this ansatz into \eqref{appeq:fpt_laplace} enforces
\begin{align}
	\omega_{1,2} = - \frac{v}{2D_x} \pm \frac{\sqrt{v^2 + 4 D_x s}}{2D_x}
	\label{}
\end{align}
The boundary conditions \eqref{appeq:fpt_laplace} fix the normalising constants $A, B$ to  
\begin{align}
A&=-e^{\pi  r \omega_1-\omega_1 x_1} \frac{\sinh (\pi  r \omega_2) }{\sinh(\pi  r (\omega_1-\omega_2))}\\
B&=\sinh (\pi  r \omega_1) \frac{e^{\pi  r \omega_2-\omega_2 x_1} }{\sinh (\pi  r (\omega_1-\omega_2))}
\end{align}
After some further simplifications one arrives at the $v$-dependent moment generating function
\begin{align}
		M_{x_0, x_1 }( s; v)	
	=\frac{e^{ \frac{rv}{2D_x} \left( \theta - 2\pi \right) } \sinh\left( \frac{r\sqrt{v^2+4D_x s}}{2D_x} \theta \right) -e^{ \frac{rv}{2D_x}  \theta} \sinh\left( \frac{r\sqrt{v^2+4D_x s}}{2D_x} (\theta-2\pi) \right)}{\sinh\left( \pi\frac{r\sqrt{v^2 + 4D_x s}}{D_x} \right)}
	\label{eq:quenched_bmou_MGF_exact}
\end{align}
with $\theta = (x_1 - x_0)/r$ following the notation from the main text. We note that for $v \to 0$,
\begin{align}
	\frac{r \sqrt{v^2 + 4 D_x s}}{2D_{x}} \to \sqrt{\bar{s}}
	\label{}
\end{align}
and one recovers the undriven moment generating function 
\begin{align}
		M_{x_0, x_1}( s; v=0) =M_{x_0, x_1 }( s) = \frac{\sinh\left( \theta \sqrt{\bar{s}} \right) - \sinh\left( (\theta - 2 \pi) \sqrt{\bar{s}} \right)}{\sinh\left( 2\pi \sqrt{\bar{s}} \right)}= \frac{\cosh\left( (\theta - \pi)\sqrt{\bar{s}} \right)}{\cosh\left( \pi \sqrt{\bar{s}} \right)}
	\label{}
\end{align}
in agreement with the independently found expression \eqref{eq:bm_classic_mgf}. In App.~\ref{app:equivalent_quenched_averages}, we show that to second order in $v$ this result is identical to the first-order correction $M^1$ from Eq.~\eqref{eq:bmou_full_mgf} in the limit of $\beta \to 0$.

\section{Equivalence of quenched averages}
\label{app:equivalent_quenched_averages}

In  this section, we provide  a more detailed proof showing that Eq.~\eqref{eq:quenched_mgf_are_the_same} indeed holds for the case of Brownian Motion driven by coloured noise, \ie that our framework perturbatively gives the correct moment generating function of first-passage times when taking the quenched average over Eq.~\eqref{eq:quenched_langevin} with normal distributed drift $v \sim \mathcal{N}(0, \varepsilon^2 w^2)$. 
To that end, we take the $\beta \to 0$ limit of the analytically found $M^1$  (\cf Eq.~\eqref{eq:mgf1_bmou}),
\begin{align}
\label{appeq:quenched_limit_mgf1_bmou}
\lim_{\beta \to 0} M^1 = &\frac{1}{8  \sqrt{\bar{s} }} \Big\lbrace \theta  \cosh \left(\theta  \sqrt{\bar{s} }\right) \left(\theta  \sqrt{\bar{s} }-\tanh \left(\pi  \sqrt{\bar{s} }\right)\right) \\
&+\sinh \left(\theta  \sqrt{\bar{s} }\right) \left[-\left(\theta ^2-2 \pi  \theta +2 \pi ^2\right) \sqrt{\bar{s} } \tanh \left(\pi  \sqrt{\bar{s} }\right)+2 \pi  (\pi -\theta ) \sqrt{\bar{s} } \coth \left(\pi  \sqrt{\bar{s} }\right)+\pi  \tanh ^2\left(\pi  \sqrt{\bar{s} }\right)+\theta -\pi \right]\Big\rbrace. \nonumber 
\end{align}
In Eq.~\eqref{eq:quenched_mgf_are_the_same}, it is claimed that this equals
\begin{align}
   \frac{D_x \alpha}{2\eps^2 w^2} \int_{-\infty}^{\infty} \dd{v} \frac{e^{-\frac{v^2}{2\eps^2 w^2}}}{\sqrt{2\pi \eps^2 w^2}} v^2 \left. \partial_v^2 \right|_{v=0} M_{x_0, x_1}( s; v) \ .
   \label{appeq:derivative_mgfv}
\end{align}
Evaluating this expression using the result from \Eqref{eq:quenched_bmou_MGF_exact} and setting $v=0$  results in
\begin{align}
\label{appeq:second_derivative}
    \frac{D_x \alpha}{2}  \left. \partial_v^2 \right|_{v=0}M_{x_0, x_1}( s; v) =& \frac{1}{8 \sqrt{\bar{s} } } \Big\lbrace \frac{1}{\text{sinh}   \left(2 \pi  \sqrt{\bar{s} }\right)} \Big[\sinh \left((\theta -2 \pi ) \sqrt{\bar{s} }\right) \left(2 \pi  \coth \left(2 \pi  \sqrt{\bar{s} }\right)-\theta ^2 \sqrt{\bar{s} }\right) +\theta  \cosh \left(\theta  \sqrt{\bar{s} }\right) \nonumber \\
    &+(2 \pi -\theta ) \cosh \left((\theta -2 \pi ) \sqrt{\bar{s} }\right)+\sinh \left(\theta  \sqrt{\bar{s} }\right) \left((\theta -2 \pi )^2 \sqrt{\bar{s} }-2 \pi  \coth \left(2 \pi  \sqrt{\bar{s} }\right)\right)\Big] \Big\rbrace\ . \nonumber
\end{align}
The expressions in \eqref{appeq:quenched_limit_mgf1_bmou} and \eqref{appeq:second_derivative} are indeed equal as can be shown with the help of Mathematica \cite{Mathematica}.

\section{Explicit expressions for functional derivatives of transition probability densities of Ornstein-Uhlenbeck processes}
\label{app:explicit_densities_ou}
Following the notation from Sec.~\ref{subsec:OUOU_eigenfunctions}, we here give the explicit expressions of $\hat{T}^{(1)}, \hat{T}^{(2)}$ as implicitly given in Eqs.~\eqref{eq:ouou_rho1xy}-\eqref{eq:ouou_rho2xy}. We confine ourselves to the case of $\start < \target$. Starting from the Fourier-transformed transition probability density (\cf \Eqref{eq:ouou_rho0xy}), all other functional derivatives are given as partial derivatives of this density. From formula \eqref{eq:ouou_rho1xy}, one obtains
    \begin{align}
\transcoeff{1}{-i\bar{s}\alpha, -i\beta}  = -\frac{e^{\frac{{\start}^2-{\target}^2}{4} }}{\sqrt{2 \pi }\ell^2 \alpha^2(1-\bar{\beta}) } \left[ \Gamma (\bar{\beta} +\bar{s} ) D_{-\bar{\beta} -\bar{s} }(-\start) D_{-\bar{\beta} -\bar{s} +1}(\target)-\Gamma (\bar{s} +1) D_{-\bar{s} -1}(-\start) D_{-\bar{s} }(\target)\right]
\end{align}
where $\Gamma(\bar{s})$ is the usual Gamma-function. Letting $\start \to \target$, gives the first functional derivative of the return probability at $\target$,
\begin{align}
\returncoeff{1}{-i\bar{s}  \alpha, -i\beta} = -\frac{\Gamma (\bar{\beta} +\bar{s} ) D_{-\bar{\beta} -\bar{s} }\left(-\target\right) D_{-\bar{\beta} -\bar{s} +1}\left(\target\right)-\Gamma (\bar{s} +1) D_{-\bar{s} -1}\left(-\target\right) D_{-\bar{s} }\left(\target\right)}{\sqrt{2 \pi }\ell^2 \alpha^2(1-\bar{\beta}) }\end{align}
These results further imply
\begin{align}
\transcoeff{1}{-i\bar{s} \alpha  -i\beta, i\beta}  =- \frac{e^{\frac{\start^2-\target^2}{4}} (\Gamma (\bar{s} ) D_{-\bar{s} }(-\start) D_{1-\bar{s} }(\target)-\Gamma (\bar{\beta} +\bar{s} +1) D_{-\bar{\beta} -\bar{s} -1}(-\start) D_{-\bar{\beta} -\bar{s} }(\target))}{\sqrt{2 \pi } (1+\bar{\beta} )\ell^2\alpha^2}\end{align}
and
\begin{align}
\returncoeff{1}{-i\bar{s} \alpha -i\beta, i\beta}  = -
\frac{\Gamma (\bar{s} ) D_{-\bar{s} }\left(-\target\right) D_{1-\bar{s} }\left(\target\right) -\Gamma (\bar{\beta} +\bar{s} +1) D_{-\bar{\beta} -\bar{s} -1}\left(-\target\right) D_{-\bar{\beta} -\bar{s} }\left(\target\right)}{\sqrt{2 \pi } (1+\bar{\beta} )\ell^2 \alpha^2}\end{align}
The second order derivative of the transition probability is
\begin{align}
\transcoeff{2}{-i\bar{s} \alpha, -i\beta,i\beta}  =\frac{e^{\frac{\start^2-\target^2}{4}}}{ \sqrt{2 \pi } \left({\bar{\beta}} ^2-1\right)\ell^3 \alpha^3} 
&\left[ 2 \Gamma (\bar{\beta} +\bar{s} +1) D_{-\bar{\beta} -\bar{s} -1}(-\start) D_{-\bar{\beta} -\bar{s} +1}(\target) \right. \nonumber \\
& \left.+(\bar{\beta} -1) \Gamma (\bar{s} ) D_{-\bar{s} }(-\start) D_{2-\bar{s} }(\target)-(\bar{\beta} +1) \Gamma (\bar{s} +2) D_{-\bar{s} -2}(-\start) D_{-\bar{s} }(\target)) \right]
\end{align}
and of the return probability
\begin{align}
\returncoeff{2}{-i\bar{s} \alpha, -i \beta, i\beta} =\frac{1}{ \sqrt{2 \pi } \left({\bar{\beta}} ^2-1\right)\ell^3 \alpha^3} 
&\left[ 2 \Gamma (\bar{\beta} +\bar{s} +1) D_{-\bar{\beta} -\bar{s} -1}(-\target) D_{-\bar{\beta} -\bar{s} +1}(\target) \right. \nonumber \\
& \left.+(\bar{\beta} -1) \Gamma (\bar{s} ) D_{-\bar{s} }(-\target) D_{2-\bar{s} }(\target)-(\bar{\beta} +1) \Gamma (\bar{s} +2) D_{-\bar{s} -2}(-\target) D_{-\bar{s} }(\target) \right]
\end{align}

\end{widetext}

\end{document}